# Decentralized lending and its users: Insights from Compound

Kanis Saengchote*

*Chulalongkorn Business School*

This version: 12 December 2022

***ABSTRACT***

Permissionless blockchains offer an information environment where users can interact privately without fear of censorship. Financial services can be programmatically coded via smart contracts to automate transactions without the need for human intervention or knowing users' identity. This new paradigm is known as decentralized finance (DeFi). We investigate Compound – a leading DeFi lending protocol – to show how it works in this novel information environment, who its users are, and what factors determine their participation. On-chain transaction data shows that loan durations are short (31 days on average), and many users borrow to support leveraged investment strategies (yield farming). We show that systemic risk in DeFi arises from concentration and interconnection, and how traditional risk management practices can be challenging for DeFi.

* Corresponding author. Chulalongkorn Business School, Chulalongkorn University, Phayathai Road, Pathumwan, Bangkok 10330, Thailand. (email: kanis@cbs.chula.ac.th). I would like to thank Unnawut Leepaisalsuwanna for helping me understand how smart contracts and blockchains work, Carlos Castro-Iragorri for helping me understand Compound's data API, and Supakorn Phattanawasin for excellent research assistance. I am grateful for the comments from the editor and the four anonymous referees who have been immensely helpful in reshaping and sharpening this paper. All remaining errors are my own.

0

# 1. Introduction

The term decentralized finance (abbreviated as "DeFi") carries a very a specific meaning: it refers to an alternative financial system built on a permissionless blockchain that promises openness, efficiency, transparency, interoperability, and decentralization (Harvey et al., 2021; Schär, 2021), where users can anonymously interact in a cash-like digital environment (Nakamoto, 2008). DeFi involves a system of computer algorithms, often called a "DApp" (short for decentralized application) or "protocol", that replicates traditional financial services such as lending, exchange, and asset management. A protocol is structured via a series of smart contracts, some of which represent transferable data also known as "tokens". Smart contracts and tokens can be connected to form a network across any number of protocols without the need to ask for permission to interact with, so developers can freely incorporate existing information in the blockchain.[1] This interoperability is often referred to as "composability", and DeFi money is referred to as "Lego money" for this reason.

In its simplest abstraction, traditional lenders raise funds in various forms: deposits, bills of exchanges, or shareholder equity, and lend them out to those who need capital in exchange for interest income. In doing so, they take on several risks, such as interest rate risk (fixed versus variable rates), asset-liability mismatch risk (market value and maturity) and credit risk (ability and willingness to repay). In other words, lenders not only connect but transform the needs of suppliers and borrowers of capital in a way that an agreement can be reached, take on the task of monitoring borrowers (Diamond, 1984), and create liquidity in the form of deposits (Gorton and Pennacchi, 1990; Kashyap et al., 2002).

In DeFi, where there is no such centralized institution to take on these risks and identities of participants are unknown due to optional anonymity, mechanisms need to be designed so that suppliers and borrowers of capital can still reach an agreement, and this is the role of the smart contracts in a DeFi lending protocol. Thus, "decentralized lending" in practice means lending in the decentralized information environment of a permissionless blockchain. In an environment

---

[1] This freedom is not absolute, however, as owners of smart contracts that crate the tokens can program restrictions, prohibiting certain addresses from interacting with the contract, effecting banning transfers. An example of this is the Tether USD (USDT) stablecoin, which contains an "addBlackList" function. Thus, the degree of freedom depends on the standpoint of the developer. See
https://etherscan.io/address/0xdac17f958d2ee523a2206206994597c13d831ec7#writeContract for details of the function.

where anyone can anonymously and freely participate, interesting phenomena such as creating new accounts to receive rewards ("airdrop") can occur, and it leads to the inability to enforce certain types of risk management practices such as exposure limits to avoid concentration risk.

In this paper, we show that Compound – one of the earliest and largest lending protocol built on the Ethereum blockchain – does not take on interest rate risk, asset-liability mismatch risk, faces very small credit risk, and does not require any external funding. In fact, Compound could be considered a mutual lender, where depositors mutually own the economic benefits to the protocol, like how a mutual insurance company operates. This allows Compound to intermediate funds in the pseudonymous information environment.

Between May 2019 and June 2020, Compound has supplied more than $61.1 billion in token loans to almost 23,000 borrowers. Depositors and borrowers in Compound are concentrated, as the top 100 addresses account for 75% of all deposits and the top 100 borrower addresses 78% of all loans. This concentration is greater than a traditional bank. For example, Juelsrud (2021) finds that the top 5% of depositors in Norway in 2018 accounted for 53% of all deposits. We document the factors that influence aggregate depositing and borrowing activities, particularly how reward distribution can affect users' incentives to borrow, such as effectively making total return from collateralized borrowing positive even after accounting for borrowing cost.

Because DeFi protocols do not restrict the use of proceeds, loans can be used to finance spending and capital investments, or fund leveraged investment positions. Using DeFi protocols to maximize rewards has come to be known as "yield farming, which can be amplified by leverage, resulting in "leveraged yield farming" investment strategies. With address-level and loan-level data retrieved directly from the blockchain, we find evidence that are more consistent with leveraged investments rather than financing, such as shorter durations for certain types of addresses and recursive interactions with the protocol to maximize rewards.

Further investigations of address-level activities and new information on the identity of the addresses' owners who are high-profile, failed crypto businesses such as Three Arrows Capital, Celsius Network, Alameda Research, and FTX, we find that on-chain transactions before the collapse of the crypto asset market in May 2022 can reveal potential sources of vulnerabilities that built up as systemic risk and threatened the crypto asset market.[2] As DeFi loans involve automatic liquidation when loans are insufficiently collateralized, Aramonte et al. (2021) show that following forced DeFi liquidations, crypto asset prices fall sharply and volatility spikes follow. Given that leverage is procyclical, liquidation risk can amplify instability of the DeFi ecosystem. In addition, Acemoglu et al. (2015) show that dense interconnections can propagate shocks rather than enhance financial stability, and on-chain data shows that this is indeed the case for DeFi. The systematic risk arising from interconnections has been highlighted by many, including The Financial Stability Oversight Council (FSOC),[3] and lending protocols can contribute to such risk.

As DeFi is an emergent field, there is little extant research on the issue, particularly on lending. Several papers explain how DeFi lending protocols and crypto shadow banking work (Bartoletti et al., 2020; Gudgeon et al., 2020; Perez et al., 2020; Kozhan and Viswanath-Natraj, 2021; Li and Mayer, 2021; Castro-Iragorri et al., 2022), with Perez et al. (2020) and Castro-Iragorri et al. (2022) specifically investigating Compound. However, most of the papers approach the issue at a conceptual level or rely on aggregate flow data. In contrast, our paper uses transaction-level blockchain data to provide a more microscopic view on the issue.

The rest of this paper is organized as follows. Section 2 provides an overview of what DeFi means and how Compound works. Section 3 outlines data sources, research hypotheses, and empirical methodologies. We present the results of aggregate-level and address-level analyses of Compound's usage in Section 4 and conduct further investigations into concentration, connectivity, and systemic risk in the context of the crypto asset market collapse in 2022 in Section 5.

## 2. How Compound Works

### 2.1 Properties of Decentralized Finance

In this section, we describe the properties and operational rules of DeFi so readers can understand the context and why financial services offered on permissionless blockchains differ from traditional finance. While there had been many interpretations of what the "decentralized" part of DeFi means, by 2022, the general definition has converged toward the ability offer financial products and services without relying on a "trusted central intermediary" such as a bank or a payment processor (Aramonte et al., 2022; Carapella et al., 2022). Financial records including money can be represented by data, and such central intermediary who is the gatekeeper to the information system may abuse this database privilege, so an alternative environment where users are not subjected to this authority may be desired.

---

[2] Source: https://www.nytimes.com/2022/05/12/technology/cryptocurrencies-crash-bitcoin.html, accessed November 28, 2022.
[3] Source: https://home.treasury.gov/system/files/261/FSOC-Digital-Assets-Report-2022.pdf, accessed November 28, 2022.

This definition is consistent with the objective of the Bitcoin white paper, where Nakamoto (2008) expresses a desire of building "a peer-to-peer electronic cash system" to preserve the privacy and convenience of money that physical cash provides in the context of an increasingly digitized world. Such system would allow users to remain anonymous and retain the ability to transact without censorship, which requires a different information environment compared to the ones used by the financial system leading up to 2008. Blockchain technology is one component required to ensure that users of such system accept the states and flows of information recorded, obviating the need for a centralized, trusted third party as gatekeeper of the system. This interpretation is consistent with the view of Cong and He (2019) who describe blockchain technology as providing "decentralized consensus".

A blockchain with decentralized consensus allows potentially any entity running the blockchain software (a "node") to hold the authority to record new data as a block, to be added to the pre-existing blocks of data, and other nodes would then agree and update their information (or ledger) accordingly. As each node records the same version of the ledger, blockchain technology is often referred to as distributed ledger technology (DLT). The node chosen by the consensus algorithm (for example, proof-of-work or proof-of-stake) is rewarded with data known as "native coins" (such as Bitcoin) as compensation for their efforts.

The openness of this information environment entails both the transparency of information recorded on the blockchain and the freedom to interact with other users on the same blockchain, leading to adjectives such as "permissionless" and "censorship-resistant" being associated with this environment. The censorship resistance property of the blockchain depends on how decentralized the nodes are, and the amount of data recorded in each block is constrained by the minimum capability of an entity who can become a node in this environment; in other words, the more decentralized the environment is, the less efficient the blockchain would be. Block space is a scare resource in a permissionless blockchain network.

The early blockchains have limited capabilities and permissible data is restricted, as they were built for the specific task of internal remittances and self-custody of funds.[4] For example, the Bitcoin blockchain and its variants only allow existing Bitcoin (BTC) to be transferred between accounts. The accounts are known as "addresses," because users can freely create them with no personally identifiable information required, and thus addresses are merely destinations to send the coins to, not persons. This objectification of data where ownership is proven by possession (like bearer instrument) rather than identify of owner is the distinction between token money and account-based money described by Brunnermeier et al. (2019).

Later versions of blockchains allow more information to be recorded, including programming codes which would then be executed by the network of computers running the blockchain software. These blockchains have come to be known as "programmable blockchain", and the most prominent example is the Ethereum blockchain. In programmable blockchains, users are required to pay transaction fee using the native coins (so, Ether is used to pay for transactions in the Ethereum blockchain) to incentivize nodes to compute and record new information on to the blockchain because decentralized computing power and block space are limited resources in such blockchains. This transaction fee is known as "gas", as it is needed to incentivize the operation of the programmable blockchain network.

When addresses contain codes, they are referred to as "smart contracts" which can be used to compute and record new data on to the blockchain, thus allowing a wider range of possibilities beyond transfers and payments. Smart contracts can also create new numerical data and design how users can interact with them; such data is referred to as "tokens", while "coins" are often used to describe numerical data created natively by the blockchain software, hence native coins.

While in principle, addresses can interact directly in a peer-to-peer manner, their anonymity makes communication and negotiation difficult and can introduce counterparty risk for intertemporal transactions. Address owners can voluntarily identify themselves or be flagged by others, hence addresses are often described as being "pseudonymous" rather than anonymous, but new addresses can always be freely created. Thus, for agreements to be enforceable, DeFi service providers (referred to as "protocols") would create smart contracts to hold tokens that are relevant for a transaction as a pool to reduce failure to deliver (FTD).

---

[4] For more on limitations of such early blockchains, see John et al. (2022).

Because of blockchain transparency, users can see the availability of tokens in the pool and the computing network can check whether a desired transaction can be performed, mathematical functions can be written so that users can interact with the pool to exchange or borrow tokens, depending on the purpose of the protocol. The outcome of the mathematical calculations such as exchange rate and interest rate are updated and recorded every time a new block is added to the blockchain. Details of how the main types of DeFi protocols operate are provided in the Appendix.

In sum, smart contracts created as part of protocols by developers become a new type of intermediary, operating autonomously as instructed by computer codes rather than by human discretion. Therefore, decentralization in the context of DeFi thus means financial intermediation via smart contract pools in a permissionless (decentralized) blockchain.

## 2.2 An Overview of Compound

Founded in 2017, Compound launched its "money market" protocol in September 2018. The protocol's mission is to generate an efficient system for earning interest, which is achieved by a dynamic interest rate algorithm that automatically adjusts borrowing and saving rates as a function of available liquidity. Because it needs to generate token income to pay depositors, Compound can also be viewed as a lending protocol, but in many ways, it seems to prioritize depositors over borrowers. Users must deposit accepted tokens into Compound's cToken contracts, which return the cToken as depository receipts. For example, if a user deposits DAI, she would receive cDAI token in return. The cToken is often referred to as "wrapped" token, as it is created by wrapping DAI by a cDAI contract to create a linked token.

The cToken contracts set the exchange rates between the tokens according to accrued saving rates, giving users more of the deposited tokens when redeemed. This design means interest payment is made upon redemption like discount loans, so users will pay gas cost only when they decide to withdraw. To see how this works let us consider an example. When a user first deposited DAI to the cDAI contract, the exchange rate may be 46.2896 cDAI to 1 DAI. One day later, the exchange rate may move to 46.2859, so 1 DAI deposited will now be redeemed for 46.2896 / 46.2859 = 1.0000081 DAI after one day, equivalent to 3% annually. The flexible interest rate is set automatically via computer code, to be described later. Figure 1 provides a high-level overview of how the cDAI contract operates through the lens of a traditional balance sheet.

**Figure 1: Balance sheet view of Compound cToken contract.**
DeFi protocols can be represented as a business entity with assets and liabilities. The solid lines represent tokens held in address, while dotted lines represent financial relationships between addresses. Tokens are transferable objects in blockchains: if an address holds some tokens, they are the address' assets. However, other assets such as loans are financial contracts representing the repayment obligations of the borrowers. In this case, the borrowed tokens will be transferred to the borrower's address, but the lender and borrower are tied by a promise which may or may not be tokenized. Similarly, when a user deposits DAI, the tokens will be held by the cDAI contract. But the cDAI contract will issue cDAI tokens as depository receipts, which are held by the depositor. In this sense, the depositor can also be considered a creditor of the cToken contract, so the cDAI token could be considered tokenized debt. Like how bank deposits are considered as debt to the bank, the cDAI token can also be classified as tokenized deposit. Details of the contract can be viewed at https://etherscan.io/address/0x5d3a536E4D6DbD6114cc1Ead35777bAB948E3643.

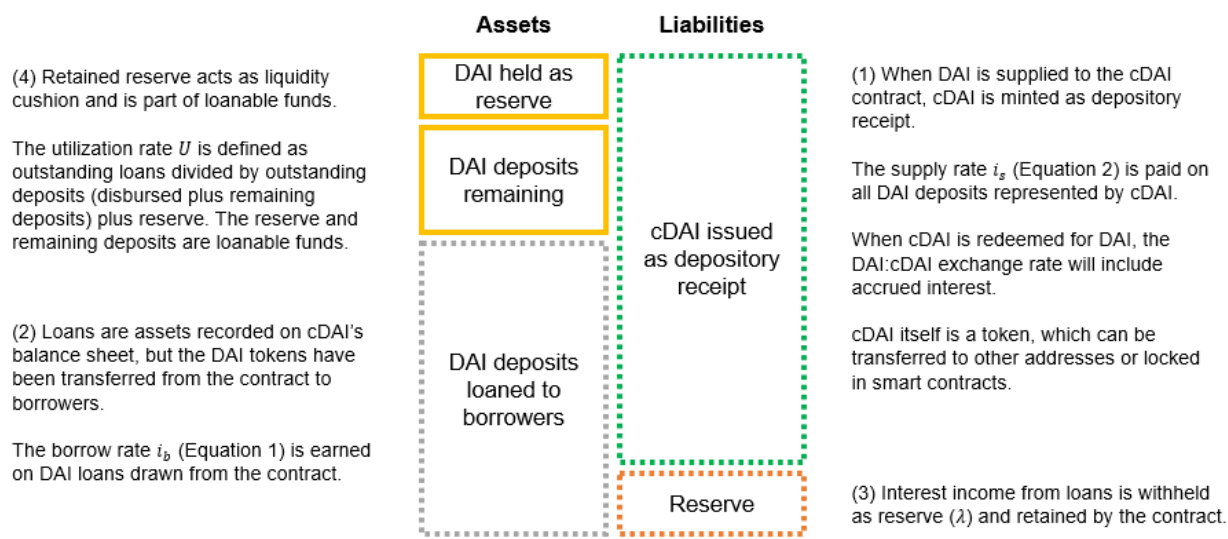

When it began, Compound initially accepted 4 tokens, where were Ether (ETH), 0x Protocol (ZRX), Basic Attention Token (BAT), and Augur (REP). As of July 2021, the list stood at 12 tokens, including stablecoins such as DAI, USD Coin (USDC) and Tether (USDT), and its governance token, COMP, whose properties will be described later. Stablecoins are tokens designed with mechanisms to stabilize their exchange rate to a pre-determined value, such as national currency, providing users with lower exchange risk compared to other crypto assets.[5] As of July 26, 2021, the most popular deposited tokens in Compound were USDC ($5 billion), followed by DAI ($4.3 billion) and ETH ($3.3 billion).[6] Figure 2 Panel A shows the daily value of

---

[5] See Klages-Mundt et al. (2020) for the different mechanisms behind the creation and price stabilization of stablecoins.
[6] Source: https://compound.finance/markets, accessed on July 26, 2021.

outstanding token deposits against ETH price. While most of dollar value of deposits is in stablecoins, net outstanding value tracks movements in ETH price well.

The deposited tokens become part of the liquidity pool that can then be loaned out to users. Like its traditional money market fund counterpart, Compound follows the Kashyap et al. (2002) intermediation banking model, where loanable funds are deposited tokens minus loaned tokens and does not create new money like how modern banks operate (McLeay et al., 2014). Users who want to borrow must first deposit accepted tokens as collateral and maintain sufficient overcollateralization (which is the inverse of the loan-to-value ratio) or face liquidation. This makes DeFi loan more like a repurchase agreement than a credit agreement. With the pseudonymity of participants in the ecosystem, lenders have limited ability to ensure that borrowers honor their obligations. To reduce asset-liability mismatch risk, Compound require borrowers to pledge crypto assets which are valuable, easily repossessed, and liquid.

To ensure that repossession can be automated and constitute repayment, the collateral must reside on the same blockchain for smart contracts to have any authority over them (hence, they must be tokens) and be easily resaleable (that is, they must have liquid markets). This restricts the space of permissible collaterals. Tokens that represent claims on off-chain assets such as tokenized real-world assets (for example, real estate or vehicles) present addition enforcement costs and counterparty risk, or heterogenous tokens that represent unique claims such as virtual assets represented by non-fungible tokens, or fungible tokens which are inactively traded or are traded in limited venues are examples of less desirable collateral. Corporate finance research shows that collateral quality can influence debt contracts in various ways. For example, Benmelech and Bergman (2009) show that collateral redeployability can affect credit spreads and loan-to-value ratios. For Compound, differences in collateral quality are reflected solely in the loan-to-value ratios, and not all crypto assets are accepted as collateral.

Figure 2 Panel B shows the dollar value of token loans originated by Compound in each month. While loans can be drawn in any token that Compound accepts as collateral, stablecoins still form the majority of loans in the protocol, with DAI the most popular, followed by USDC.

**Figure 2: Compound activities.**
Panel A plots the daily dollar value of outstanding token deposits between May 2019 and June 2021 and daily ETH price (right-hand side scale). The top-five tokens are ETH, WBTC (wrapped Bitcoin), DAI, USDC and USDT. Panel B plots the dollar value of monthly token loans drawn from cToken contracts during the same period.

Panel A: Daily net cToken outstanding in USD million

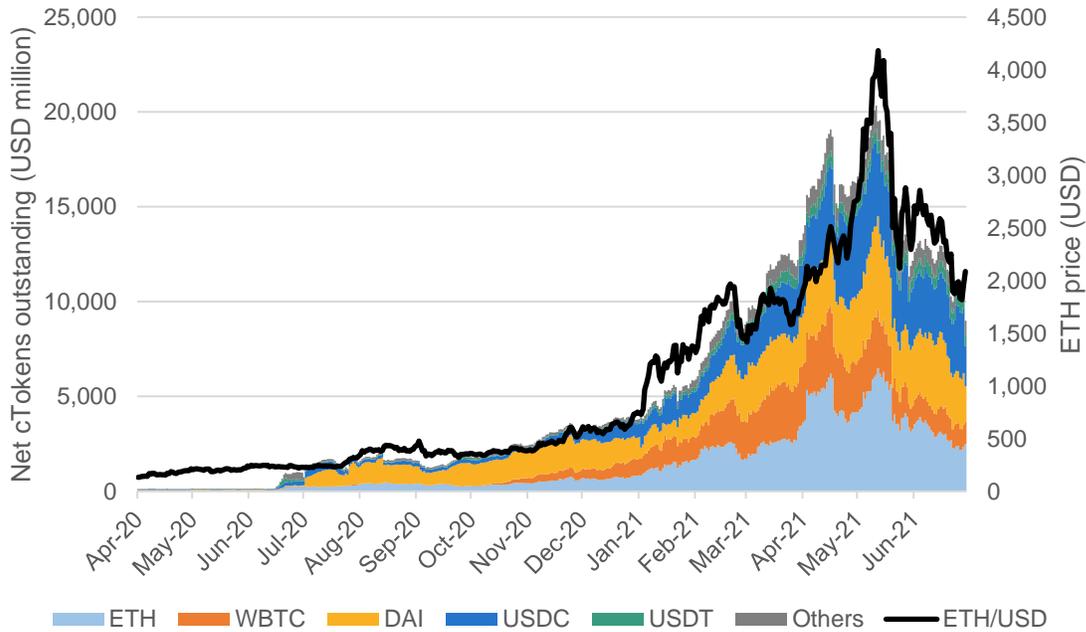

Panel B: Monthly token loans originated in USD million

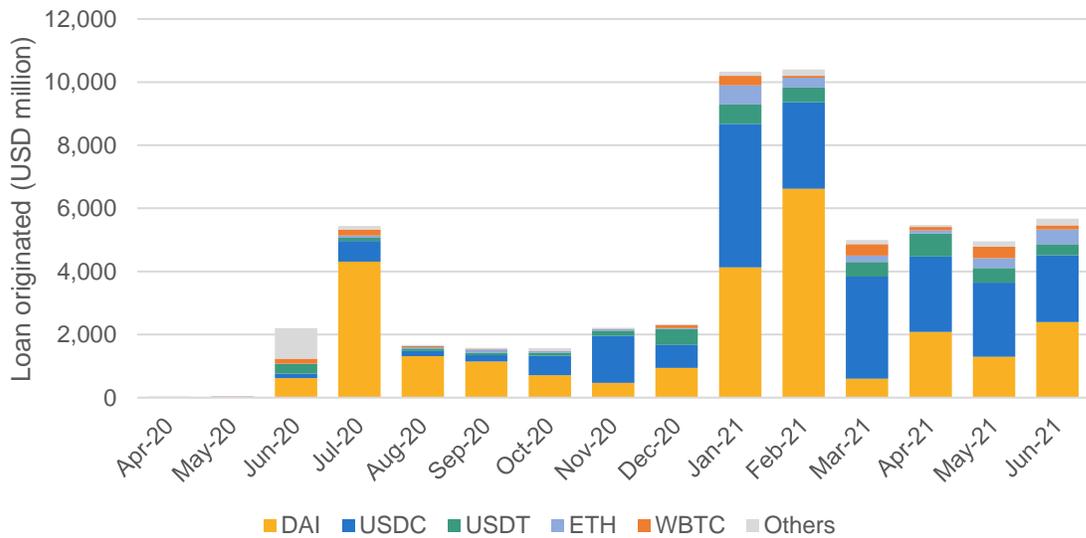

A governance token is a special case of a token issued by the protocol's smart contract containing voting rights. These protocol-issued tokens are sometimes called native tokens, which are different from native coins issued by the blockchain's software to incentivize the blockchain network's operation. The governance token of Compound is known as COMP, and the contract was created on March 4, 2020, with fixed supply of 10 million.[7] Holders of COMP can vote on protocol-related issues, such as adjusting interest rate models. Compound is backed by several high-profile venture capital funds, such as Andreessen Horowitz (a16z), Polychain Capital and Bain Capital Ventures,[8] who are majority holders of their governance tokens, COMP. As of July 26, 2021, the three VC firms own a combined voting power of 32.85%.[9]

Issuance of equity-like, governance tokens via initial coin offering (ICO) has received interest in the past, and several papers have argued that they can be optimal for digital platform start-ups (for example, Li and Mann, 2018; Cong et al., 2021; Gryglewicz et al., 2021; Lee et al., 2022). However, it is important to note that a smart contract is free to issue a new token without promising anything in return or does not contain any right at all, so unlike common shares trading in stock exchanges that require some degree of standardization, native tokens issued by different protocols may contain different rights.[10] While Compound's business model allows it to earn more income than expenses, the surplus is retained as reserve (see Figure 1) and there has been no explicit plan to distribute the reserves to holders of COMP. Because DeFi protocols can be freely created without filing for formal incorporation, rights and responsibilities of issuers to holders of COMP are not as well-defined as common shares or other securities.

While Compound is not the first protocol to reward its participants with its native token, it is often attributed as the force behind the "DeFi Summer" of 2020 where interest in DeFi began to pick up.[11] Within one week of its launch on June 15, 2020, the price of COMP doubled. The event is said to have kickstarted the "yield farming" phenomenon where participants interact with DeFi protocols' liquidity pools by providing them with tokens (also referred to as "staking") with expectation of native token reward in return. The reward distribution rate is often presented as annualized percentages rate and referred to as APY, thus the "yield" nomenclature was adopted. But in practice, Compound would allocate a certain quantity of COMP to a cToken contract, which would then split equally to depositors and borrowers, to be then shared on a pro-rata basis. Consequently, if there are fewer borrowers than depositors, the yield for borrowers would be higher. Compound reports this yield as APY, and users are shown these percentage rates as they interact with protocol.[12]

---

[7] The contract creation event can be viewed here: https://etherscan.io/tx/0xe87715364f1733c893b4dca5c8e932627e5ddcc2076f8fc69a9d38c5563c4ed1.
[8] Source: https://fortune.com/2019/11/14/crypto-interest-startup-compound-decentralized-finance/, accessed on October 15, 2022.
[9] Source: https://compound.finance/governance, accessed on July 26, 2021.
[10] It is unclear whether the venture capital funds invested in Compound as a protocol or as a business entity. However, ownership of COMP tokens represents protocol-specific benefits rather than claims on the company that develops the protocol.
[11] Source: https://www.coindesk.com/business/2020/10/20/with-comp-below-100-a-look-back-at-the-defi-summer-it-sparked/, accessed on October 15, 2022.

The brief DeFi Summer of 2020 that began in June ended in September, before gaining interest again at the end of 2020. As of July 26, 2021, COMP had circulating supply of 5,373,538.37 and had already distributed 973,535 with current distribution rate (also referred to as "emission" rate) of 2,312 per day.[13] As of February 20, 2022, 26.7% of supply is held in the Compound Reservoir address (treasury) and 1.39% in the Compound Comptroller address that is used to distribute COMP rewards.[14] Because of this design, we can infer the amount of COMP reward claimed by each address.

**2.3 Compound's Interest Rate Model**

While there are many mathematical equations that could be used as an interest rate model (see Gudgeon et al., 2020 for examples), the key objective for sustainable operation is to ensure that the protocol earns enough interest income from borrowers to pay depositors as expenses, and perhaps keep some as reserves for default risk or profits. Because an interest rate is the price of money, one way to endogenously achieve market equilibrium of the supply and demand of tokens is to allow prices to move freely to clear the market. Any attempt to fix prices will require some intervention.

In DeFi, deposit rates are often referred to as supply rates, and lending rates as borrow rates. Developers can decide whether they want to have fixed or floating rates in their protocol. For the case of Compound, the developers choose variable rates for both supply and borrow, thus eliminating interest rate mismatch risk (or income-expense discrepancy), but there are protocols that offer fixed rates. For example, Aave on the Ethereum blockchain offers a choice to borrow at fixed rate, while Anchor on the Terra blockchain fixes the supply rate. Compound's choice makes borrowing a highly uncertain experience, reflecting its priority toward depositors as a money market protocol.

---

[12] See https://compound.finance/governance/comp for how reward distribution and APYs are reported.
[13] Source: https://coinmarketcap.com/currencies/compound/, https://compound.finance/governance/comp, accessed on July 26, 2021.
[14] Source: https://etherscan.io/token/0xc00e94cb662c3520282e6f5717214004a7f26888#balances, accessed on February 20, 2022.

Let $i_b$ denote the borrow rate, $i_s$ the supply rate, $U$ the utilization rate, defined as outstanding loans divided by outstanding deposits plus reserve, and $\lambda$ the reserve factor. We can write Compound's interest rate model as follows:

$$i_b = \begin{cases} a + bU & if\ U \leq U^* \\ a + bU^* + c(U - U^*) & if\ U > U^* \end{cases} \tag{1}$$

$$i_s = i_b(1-\lambda)U \tag{2}$$

Under this specification, the borrow rate described by Equation 1 is a kinked linear function in $U$ with $a$ as the base rate. As $U$ exceeds some threshold $U^*$ representing the optimal or target utilization rate, the slope of $i_b$ with respect to $U$ changes from $b$ to $c$, a higher rate. This increased sensitivity both discourages borrowers from taking on new loans and encourages depositors to supply additional capital, ensuring that the lending operations will not halt. If there are insufficient tokens in the pool for a loan request, the transaction will fail. Because of the pseudonymity and high-speed nature of blockchain transactions, negotiation and communication between the borrower and the lender like traditional lending cannot be relied on, so a failed transaction in a system that is supposed to operate autonomously may cause confusion among users and induce a panic bank run (for example, Dijk, 2017; Kiss, Rodriguez and Rosa-Garcia, 2022). To ensure continuity, protocols tend design the mechanism to prevent such states from occurring. In the case of Compound, the automatic interest rate adjustment that attracts deposits and discourage loans provides that self-stabilizing mechanism.

The supply rate described by Equation 2 is moderated by $U$ (and hence no longer linear in $U$) to generate sufficient income to pay depositors, and the reserve factor $\lambda$ sets aside interest income as buffer for potential default risk, incentivize participants (via liquidator incentive, to be described in the next subsection), or retained as profits. As supply and demand of tokens in the pool change in each block, interest rates are adjusted accordingly.

The parameters of interest rate models can be changed over time. For example, DAI underwent 3 changes: the first time on April 7, 2020; the second time on May 2, 2020; and the third time on July 28, 2020. USDC underwent a single change on September 21, 2020, and USDT on August 21, 2020. Most of the changes would be proposed to the community and voted by on holders of COMP, reflecting the governance role that the token holders have.[15] Figure 3 plots the supply rates for DAI and USDC under several regimes marked using different colors and symbols. The distribution of data points suggests that they do indeed belong to different interest rate regimes.

**Figure 3: The kinked interest rate model.**
In each panel, the daily borrow rates for selected tokens are plotted against utilization rate as scatter plots. Utilization rate, computed as outstanding loan divided by outstanding deposit plus reserves, is plotted on the horizontal axis. Data is obtained from Compound's API. Compound uses a kinked interest rate model, where the interest rate is linear in utilization rate and the slope changes when utilization rate reaches the optimal level. The model is applied to borrow rate and the supply rate is further adjusted based on the utilization rate to ensure the cToken contract generates enough interest income to pay depositors and maintains sufficient liquidity. Compound adjusts the formula interest calculation several times during its operation. Data points corresponding to different interest rate regimes are marked with different colors. During the sample, the cDAI contract operated under 4 different interest rate models (v0 to v4), while the cUSDC contract operated under 2 models (v1 and v2).

Panel A: DAI

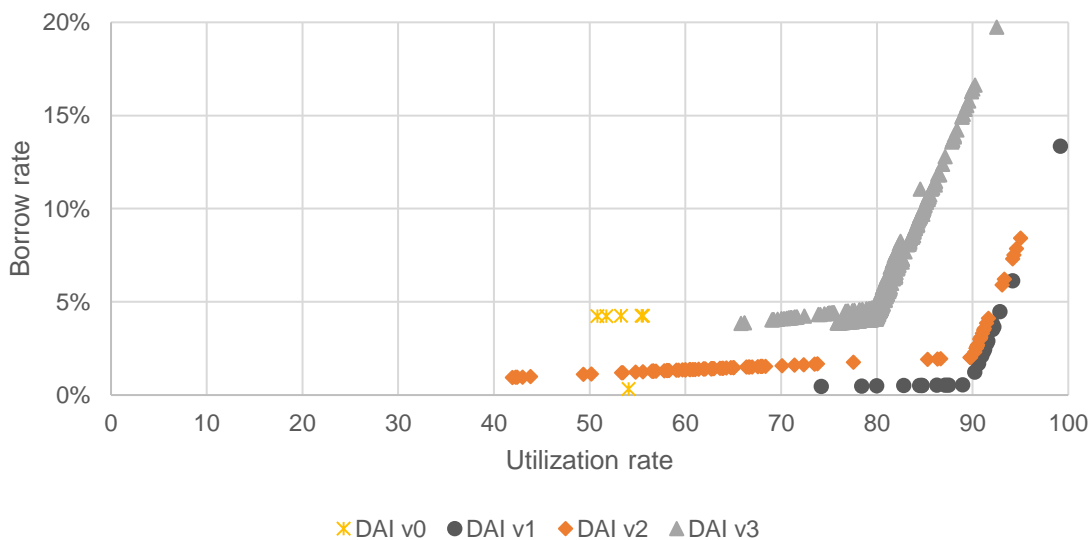

Panel B: USDC

---

[15] The first Compound governance proposal was to add USDT support. The proposal was initiated by Geoffery Hayes, the founder and CTO at Compound CTO, on April 27, 2020, and received majority support. Details of the proposal and the votes can be found here: https://compound.finance/governance/proposals/1. The second change to the DAI interest rate model change was the second proposal, proposed on April 27, 2020, and executed on May 2, 2020.

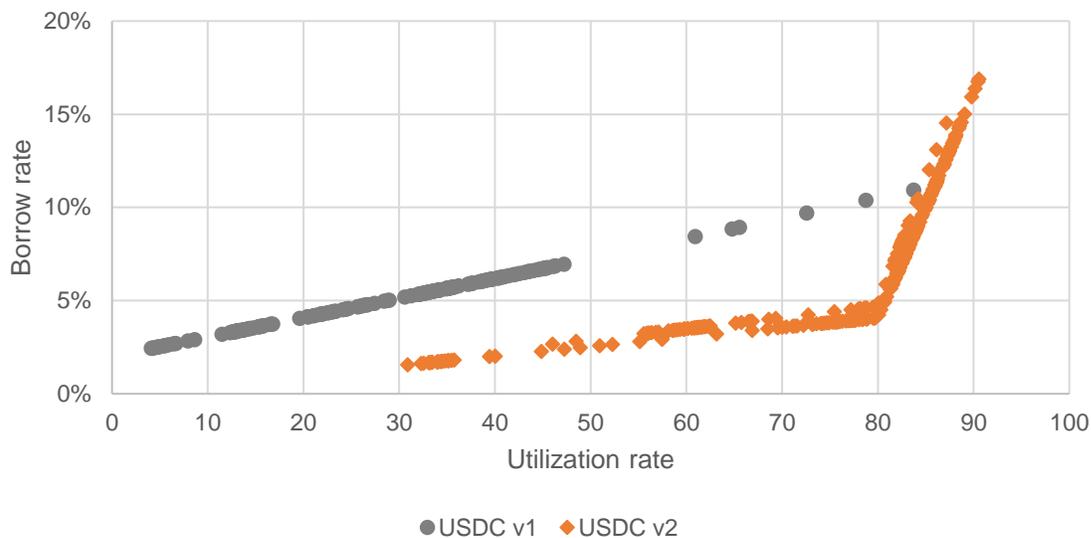

Figure 4 plots the daily rates for selected tokens computed using data obtained from Compound's API. The plots on the left show that borrow rates fluctuate with utilization rates, which is a direct consequence of the interest rate model. As Compound began distributing COMP as rewards for both suppliers and borrowers, the effective rates received and paid by users are subsidized by Compound. The plots on the right of Figure 4 display the net borrow and supply rates, computed by subtracting or adding the COMP reward rates for each activity, making supplying tokens more lucrative and borrowing less costly.

For the most part, the difference between the net supply rate and the net borrow rate are positive, making a recursive strategy of depositing tokens to borrow the same tokens, and then redepositing again profitable. This strategy will be described further in Section 2.4. The net borrow rate for tokens with low utilization rate such as ETH and WBTC are also mostly negative, as Compound disproportionately subsidizes borrowers by providing an equal number of COMP to be shared among suppliers and borrowers in the same cToken contract, so whenever there are fewer borrowers than suppliers, COMP borrow reward rates are higher.

**Figure 4: Pool interest rates and utilization rates for selected tokens.**
In each panel, the time series data of daily rates for selected tokens are plotted with utilization rate (computed as outstanding loan divided by outstanding deposit plus reserves). On the left, daily borrow rates are plotted against utilization rate. On the right, daily net borrow and net supply rates (computed by subtracting or adding the COMP reward rates for each activity) are plotted against utilization rate. Data is obtained from Compound's API.

Panel A: ETH rates

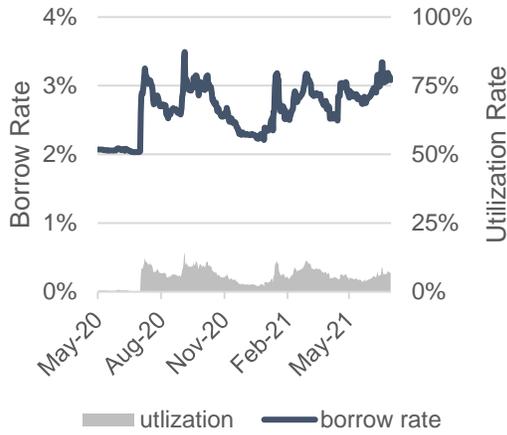
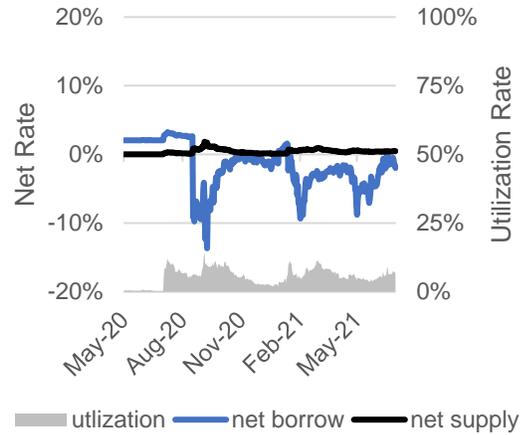

Panel B: WBTC rates

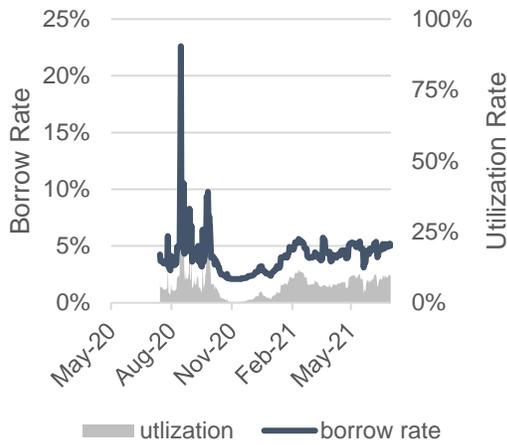
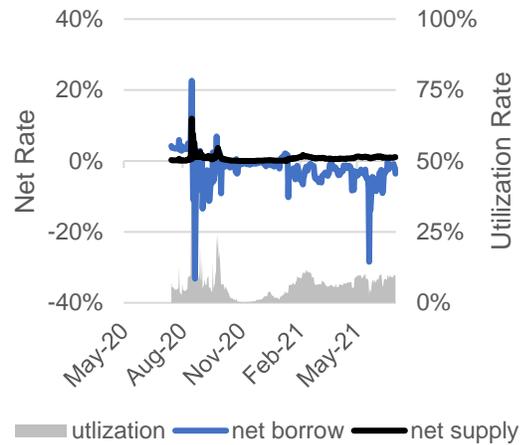

Panel C: DAI rates

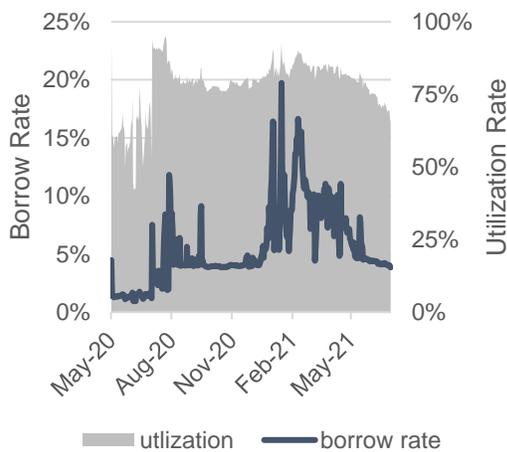
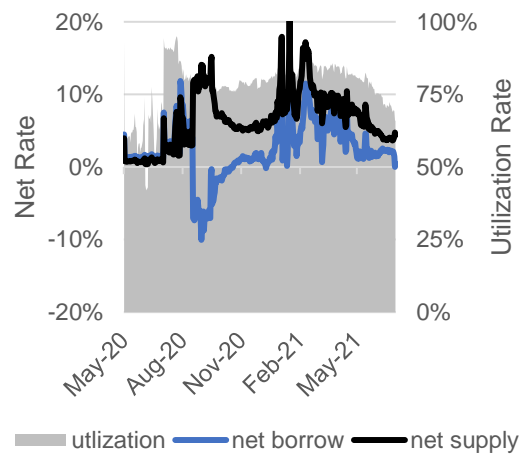

Panel D: USDC rates

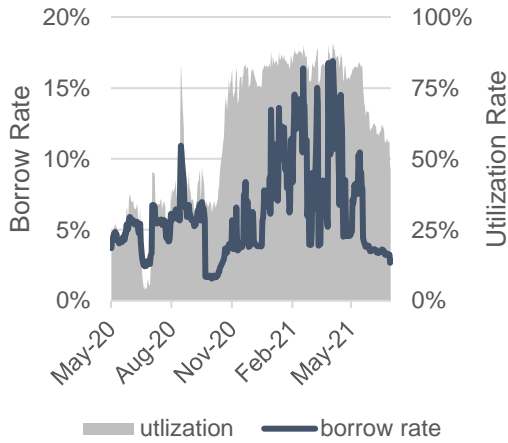
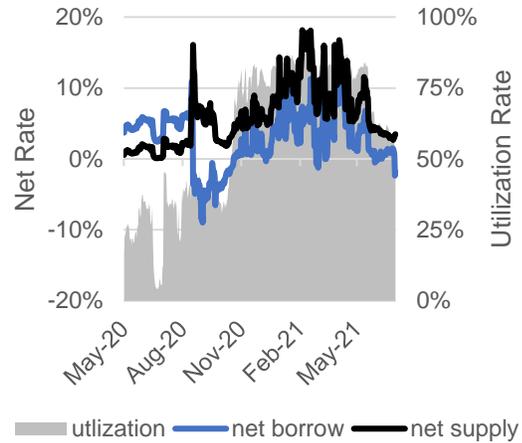

Panel E: USDT rates

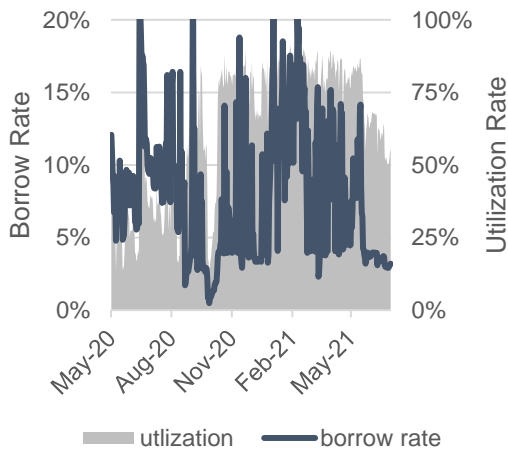
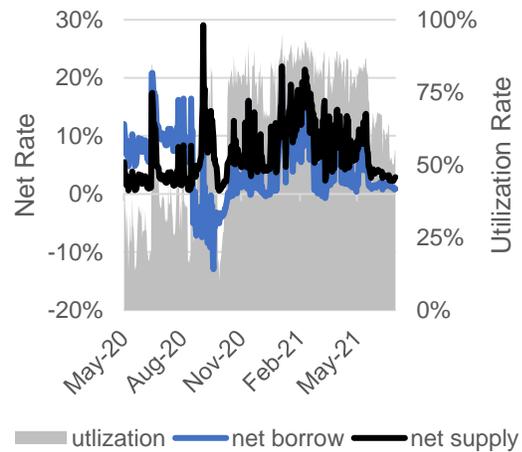

## 2.4 Compound's Loan Liquidation

Because each borrower is pseudonymous in the blockchain, credit risk assessment cannot be reliably made on information related to the borrower, so the most meaningful discriminant is the quality of the collateral. This also implies that credit risk management must rest entirely on foreclosure of collateral because the borrower is not reachable.

Compound calculates the credit line (referred to as borrow limit) of each borrower as follows. Let $(1 - \gamma_j)$ be the collateral factor for token $j$, $Q_j$ the amount of deposited token collateral and $P_j$ the price of token $j$. Token prices may vary according to trading venues, some of which are not reflected on the blockchain. To import external data, a "data oracle", a code written on the blockchain to reference external data sometimes referred to as price oracle or blockchain oracle, is required. The oracle code will specify the data source(s), and the final price used for calculation may involve processing such as averaging across sources or time.

The dollar value of outstanding loan drawn in token $i$ must be within the borrow limit $\sum_j (1 - \gamma_j) P_j Q_j$, computed across all cToken deposits, making Compound a multi-collateral protocol. In other protocols, the collateral of a debt position may be restricted to a single token only. Because scheduled payment is not required to minimize gas cost, $\gamma_j$ is set to ensure that the borrower can maintain the ability to repay, while smart contract's ability to automatically enforce transactions ensures the borrower's willingness to pay. If $P_j$ is volatile, $\gamma_j$ for that token may be set to a higher rate, reflecting the heterogeneity in collateral quality as discussed earlier. For example, as of September 13, 2021, $\gamma_{DAI} = 25\%$ (or alternatively, $(1 - \gamma_{DAI}) = 75\%$), while $\gamma_{COMP} = 40\%$.

Let $L_i$ be the quantity of loan drawn in token $i$ and $P_i$ the price of token $i$ (which could in principle be the same as some $j$; in other words, one could borrow in the same token as the collateral). At any point in time, the borrower must ensure that Equation 3 holds, and the ability to satisfy this condition depends on the relative movements between $P_i$ and $P_j$.

$$\sum_j (1 - \gamma_j) P_j Q_j - P_i L_i > 0 \qquad (3)$$

Compound refers to the expression on the left-hand side of Equation 3 as "account liquidity", which, when negative, permits third party users to partially repay the loan on the borrower's behalf and receive a share of borrower's overcollateralized tokens. In absence of gas cost, this transaction would likely be profitable, unless collateral prices change sharply over a short horizon. But because gas cost does not vary according to transaction value, it is possible that small loans are not liquidated because arbitrage profits are not sufficient, so Compound adds liquidator incentive which will be paid out of accumulated reserves to make liquidation more profitable. This is because, by design, Compound does not proactively monitor and manage credit risk of its loan positions but instead outsources the task to liquidators, probably to allow users to take part in as many aspects of the protocol as possible. Fearing liquidation, borrowers tend not to draw loans up to their full credit limits, making the loans highly overcollateralized.

Because of the collateral factors in Equation 3, liquidation is allowed when the loan is still comfortably overcollateralized. Consequently, Compound ultimately faces little to no credit risk;

only a severe drop in $P_j$ and/or a sharp rise in $P_i$ would threaten Compound with credit loss. The liquidation risk of borrowers is another aspect which makes Compound more friendly to depositors than borrowers. More technical details of how Compound's liquidation mechanism works is explained in Perez et al. (2020).

With the account liquidity in Equation 5 and the net rates example for ETH from Figure 4, it is possible for users to use a long-short strategy of the same token and never be liquidated. Consider an example where a user deposits 1 ETH as collateral, gaining maximum borrow limit of $P_{ETH}(1 - \gamma_{ETH})$. Since we are only considering ETH, we will drop the ETH subscript for brevity. The user can borrow up to $PL \leq (1 - \gamma)P$, so Equation 3 for an ETH long-short strategy after redepositing the borrowed ETH can be written as $(1 - \gamma)P + (1 - \gamma)PL - PL > 0$. Eliminating $P$, we obtain $1 - \gamma(1 + L) > 0$ which is trivially satisfied given that $L \leq (1 - \gamma)$. And if the position is liquidation-free and net borrow rate is negative, this strategy is an arbitrage. This borrow-redeposit loop can be repeated to increase leverage beyond the $\gamma$ restriction to reduce capital required to execute this strategy, making loans undercollateralized with respect to the initial ETH.[16]

Data oracles present a potential source of risk (as highlighted in the case of Iron Finance in Saengchote, 2021a), and there are service providers specialized in building trust in the data import process, but ultimately if a mistake occurs, it can be exploited, and the protocol can do little to stop it. For example, if the data oracle references a price feed from an illiquid market, a malevolent user can the manipulate the price of a token used as collateral to artificially increase the borrow limit and drain a lending pool of its loanable funds. The borrower would ultimately be liquidated, but the repossessed collateral would be worth less than the drawn loans. This vulnerability is another reason why a protocol's developer may restrict the choice of tokens permissible as collateral.

---

[16] In late 2021, Abracadabra.Money protocol was built to automate this repeated leverage process. The protocol uses interest-bearing, stablecoin-like depository receipt (similar to cDAI) as collateral to their stablecoin loans which are issued as native tokens called Magic Internet Money (MIM). This makes Abracadabra operate more like credit creation banking (McLeay et al., 2014) since MIM "deposits" are issued by the protocol, while Compound lends the tokens received as deposits, making operate more like intermediary banking with deposits as optimal security with high liquidity (Gorton and Penncchi, 1990). But while this recursive interaction can quickly increase funds locked in the protocol, it could be quick to unravel in a bank run-like fashion when trust in the collateral value is lost, as documented by Saengchote et al. (2022).

In short, Compound does not take on interest rate risk or asset-liability price and maturity mismatch risks, faces very small credit risk, and does not require external funding. In fact, in a traditional business sense, Compound's developers need not put any equity into its lending business at all, since loans are fully funded by depositors. With variable claims directly tied to the lending income, one could consider Compound a mutual lender, where depositors mutually own the economic benefits to the protocol, like how a mutual insurance company (where ownership and benefits are shared by policyholders) might operate. Depositors' contingent claim on protocol's revenue makes deposits economically very similar to equity, but without any legal rights attached to it.

In this paper, our objectives are to provide an overview of Compound's activities and their determinants, a microscopic view of who its users are, how they use the protocol, Compound's role in the DeFi ecosystem, and potential risks it may harbor. In the next section, we describe the data sources, research questions, and empirical methodologies employed.

## 3. Data and Empirical Methodology

**3.1 Data**

The Ethereum blockchain data used in this paper is obtained from Google BigQuery, which is hosted and listed on Google Cloud Marketplace. We retrieve Compound's transactions between May 2019 and June 2021, covering over 8 million deposits, 3.56 million withdrawals, 0.16 million loan draws, 0.13 million loan repayments, and 5,036 liquidations of 356,800 unique addresses. The unit of reporting in the Ethereum blockchain is an address which can be used as containers of tokens or programmed as smart contracts as described in Section 2.1. While addresses on the blockchain are in principle anonymous, owners of smart contracts typically identify themselves and provide their source codes for community audit on websites such as Etherscan.io to increase their credibility. However, this is not mandatory, and many smart contracts are unintelligible to a human reader. While the content of blockchain is transparent for all to see, all an observer can sees is binary data that cannot be easily parsed and reverse engineered.

As described earlier, blockchains allow users to interact anonymously with the information environment. Addresses are not officially labelled, so researchers must make judgement calls on how to classify addresses. In this paper, we manually inspect the content of each address to classify whether it is part of a DeFi protocol, an unidentified smart contract, or a simple address. For example, the address '0x5d3a536E4D6DbD6114cc1Ead35777bAB948E3643' is the Compound DAI (cDAI) contract that accepts DAI for cDAI from depositors and lends out DAI to borrowers.

Because of the laborious nature of the task, we only classify some of the addresses. For each transaction, we observe the source and target of token transfer, so we inspect the top 100 sources and targets of the 12 cToken contracts in terms of both token amount and frequency of transactions, resulting in 466 manually identified addresses. They are divided into 7 categories: (1) large address, (2) small address, (3) yield aggregator protocol, (4) on-ramp, (5) decentralized exchange protocol, (6) asset management protocol, and (7) unidentified contract. Large addresses are those that do not contain any programming code and do not belong to an identified protocol, while addresses and smart contracts of protocols are classified according to the criteria described in the Appendix. Other remaining addresses are classified as small addresses. We supplement the token transaction data with states of cToken contracts such as interest rates retrieved from Compound's API and token price data from CoinGecko's API. A day is defined to begin at midnight of Coordinated Universal Time (UTC), and the net deposits are defined as aggregate token flows to and from cToken contracts. The dollar value of token transactions is obtained by multiplying token quantity by the daily token prices obtained from CoinGecko's API.

Table 1 reports the aggregate summary statistics from May 2019 to June 2021 of daily net deposits defined as deposits minus withdrawals and daily token loans drawn in USD million in each of the 12 cToken contracts. DAI, USDC, USDT and TUSD are classified as stablecoins, and other tokens which are diverse in their purposes are broadly classified as cryptocurrencies for simplicity.[17] Most of the activities in both deposit and lending are in stablecoins, corresponding to the pattern observed in Figure 2. While average daily net deposits are small with medians close to zero, the maximum and minimum values can be very high. For token loans, stablecoins are also more popular, with as much as USD 3 billion stablecoin loan taken out in one day.

**Table 1: Summary statistics.**
Panel A reports the summary statistics of daily net deposits (deposits minus withdrawals) between May 2019 and June 2021 by token. During this period, Compound accepts 12 tokens. The dollar values are calculated using daily prices obtained from CoinGecko. DAI, TUSD, USDC and USDT are classified as stablecoins, and other tokens are broadly classified as cryptocurrencies (cryptos). Panel B reports the summary statistics for daily token loans drawn from the

---

[17] For example, ETH and WBTC are native coins of the Ethereum and Bitcoin blockchain, BAT and LINK are cryptocurrencies / utility tokens of Brave (web browser) and Chainlink (blockchain oracle), and COMP and UNI are governance tokens of Compound and Uniswap.

12 cToken contracts during the same period. Panel C reports the summary statistics of the supply rate, borrow rate, supply reward, borrow reward, and utilization rate.

Panel A: Daily net deposits in USD million

|  | Average | Std Dev | Min | P5 | P50 | P95 | Max |
|---|---|---|---|---|---|---|---|
| cETH | 0.00 | 0.03 | -0.12 | -0.06 | 0.00 | 0.06 | 0.12 |
| cWBTC | 0.00 | 0.00 | 0.00 | 0.00 | 0.00 | 0.00 | 0.00 |
| cBAT | 1.71 | 13.32 | -14.46 | -3.84 | 0.00 | 5.84 | 99.10 |
| cCOMP | 0.00 | 0.01 | -0.03 | -0.01 | 0.00 | 0.02 | 0.05 |
| cLINK | 0.08 | 0.26 | -0.14 | -0.11 | 0.00 | 0.67 | 1.24 |
| cREP | 0.00 | 0.05 | -0.28 | 0.00 | 0.00 | 0.00 | 0.26 |
| cUNI | 0.05 | 0.55 | -1.42 | -0.61 | -0.01 | 0.93 | 3.03 |
| cZRX | 0.11 | 1.71 | -6.19 | -1.54 | 0.00 | 1.64 | 9.76 |
| cDAI | 6.53 | 108.43 | -491.10 | -122.97 | 4.42 | 143.71 | 348.64 |
| cUSDC | 9.43 | 114.68 | -439.27 | -70.20 | 3.16 | 100.62 | 611.14 |
| cUSDT | 0.81 | 15.99 | -56.73 | -20.40 | 0.01 | 26.08 | 64.33 |
| cTUSD | 3.03 | 12.79 | 0.00 | 0.00 | 0.00 | 30.69 | 59.50 |
| All | 13.20 | 214.80 | -2,125.02 | -224.94 | 13.06 | 204.12 | 954.97 |
| Stablecoins | 12.63 | 212.91 | -2,126.06 | -219.77 | 10.30 | 195.69 | 956.14 |
| Cryptos | 0.57 | 58.30 | -990.28 | -4.58 | 0.02 | 6.66 | 537.09 |

Panel B: Daily token loans drawn in USD million

|  | Average | Std Dev | Min | P5 | P50 | P95 | Max |
|---|---|---|---|---|---|---|---|
| cETH | 5.46 | 19.00 | 0.00 | 0.01 | 0.79 | 19.00 | 200.42 |
| cWBTC | 4.41 | 15.12 | 0.00 | 0.00 | 0.22 | 20.97 | 156.48 |
| cBAT | 2.21 | 15.54 | 0.00 | 0.00 | 0.00 | 6.14 | 278.13 |
| cCOMP | 0.48 | 3.51 | 0.00 | 0.00 | 0.00 | 1.20 | 47.93 |
| cLINK | 0.20 | 1.08 | 0.00 | 0.00 | 0.00 | 0.23 | 9.63 |
| cREP | 0.64 | 6.80 | 0.00 | 0.00 | 0.00 | 0.01 | 108.86 |
| cUNI | 1.16 | 6.31 | 0.00 | 0.00 | 0.00 | 4.82 | 105.04 |
| cZRX | 0.19 | 1.78 | 0.00 | 0.00 | 0.00 | 0.26 | 29.06 |
| cDAI | 62.53 | 203.45 | 0.00 | 0.04 | 9.07 | 263.93 | 1,792.90 |
| cUSDC | 50.22 | 160.93 | 0.00 | 0.04 | 8.88 | 162.81 | 1,515.29 |
| cUSDT | 10.40 | 21.06 | 0.00 | 0.00 | 3.86 | 50.24 | 185.71 |
| cTUSD | 0.23 | 3.88 | 0.00 | 0.00 | 0.00 | 0.09 | 80.00 |
| All | 138.12 | 299.88 | 0.05 | 0.54 | 47.61 | 578.01 | 3,096.37 |
| Stablecoins | 123.37 | 294.82 | 0.02 | 0.22 | 37.81 | 553.65 | 3,066.71 |
| Cryptos | 14.75 | 35.23 | 0.00 | 0.04 | 3.45 | 82.64 | 341.39 |

Panel C: Rates and utilization in percentages

|  | Supply Rate | | Borrow Rate | | Supply Reward | | Borrow Reward | | Utilization Rate | |
|---|---|---|---|---|---|---|---|---|---|---|
|  | Average | Std Dev | Average | Std Dev | Average | Std Dev | Average | Std Dev | Average | Std Dev |
| cETH | 0.13 | 0.08 | 2.65 | 0.35 | 0.27 | 0.27 | 4.27 | 3.47 | 5.55 | 3.12 |
| cWBTC | 0.29 | 0.71 | 4.11 | 1.72 | 0.44 | 0.47 | 6.31 | 4.36 | 6.47 | 5.25 |
| cBAT | 1.46 | 3.76 | 6.83 | 6.60 | 1.30 | 1.94 | 7.96 | 22.12 | 14.28 | 19.26 |
| cCOMP | 1.98 | 3.83 | 8.69 | 5.92 | 2.35 | 1.41 | 10.31 | 5.33 | 25.43 | 10.99 |
| cLINK | 1.52 | 2.78 | 7.63 | 4.33 | 3.64 | 5.76 | 15.53 | 6.93 | 19.48 | 14.87 |
| cREP | 0.01 | 0.07 | 10.27 | 8.12 |  |  |  |  | 24.04 | 23.58 |
| cUNI | 1.24 | 2.15 | 7.53 | 4.57 | 0.77 | 0.45 | 7.04 | 4.22 | 15.76 | 13.10 |
| cZRX | 1.63 | 1.51 | 9.87 | 3.19 | 1.07 | 0.90 | 4.92 | 3.88 | 24.03 | 10.00 |
| cDAI | 3.96 | 2.50 | 5.52 | 3.34 | 2.59 | 2.26 | 3.22 | 2.83 | 78.35 | 9.03 |
| cUSDC | 3.84 | 3.08 | 6.17 | 3.18 | 2.12 | 1.61 | 3.27 | 2.90 | 60.03 | 24.80 |
| cUSDT | 4.70 | 3.94 | 7.74 | 4.49 | 2.07 | 1.86 | 2.90 | 2.65 | 64.15 | 23.41 |
| cTUSD | 0.48 | 1.02 | 1.13 | 1.35 | | | | | 19.47 | 23.18 |

Transaction level data retrieved from the Ethereum blockchain allows us to glean further insights into the behaviors of different types of users. But before providing summary statistics of the 356,800 unique addresses that interacted with Compound, we find that approximately 195,200 addresses made exactly one stablecoin deposit of 3 USD or less, with 183,000 addresses depositing exactly 3 USD. We classify these addresses as micro addresses. Figure 5 plots the numbers of new depositors and borrowers between May 2019 and June 2021. The spike in the number of new micro addresses began in October 2020, and again in December 2020, but there no noticeable change in new borrower addresses around those dates.

The micro addresses do not further interact with Compound and leave their funds deposited throughout the sample period. Our best conjecture regarding these peculiar interactions is that the addresses were created in anticipation of COMP token distribution from Compound as reward for interacting with the protocol. In September 2020, Uniswap (a decentralized exchange protocol that operates by requiring users to participate as market makers by depositing tokens into liquidity pools) began distributing UNI, its governance token. Users who provided liquidity to the protocol prior to the distribution date would receive at least 100 UNI, which was priced at USD 3.50 soon after the distribution.[18] This reward mechanism is known as an "airdrop" in the DeFi community, and the airdrop of UNI was an unanticipated but welcomed surprise for Uniswap's users.

Around the same time, users on Compound's community discussion board began discussing whether COMP should also be distributed as an airdrop,[19] and again in late November,[20] which coincides with the spike in micro addresses' activities. Because blockchain technology allows new addresses to be freely generated, these 195,200 micro addresses could belong to a much smaller number of users who essentially hold free real options on airdrops, and a stablecoin deposit would provide a downside protection of their funds. Compound has yet to distribute any airdrop, but in this context, the real option does not have a maturity date. Coupled with the rapid rise in ETH price from around 350 USD in October 2020 to more than 1,000 USD by January 2021 and more than 4,000 USD by May 2021, the gas cost of transferring the stablecoins out of the addresses likely exceeds the option value of leaving their tokens with the protocol. Figure 5 suggests that such practice was no longer popular by mid-2021.

---

[18] Source: https://www.coindesk.com/markets/2020/09/17/uniswap-recaptures-defi-buzz-with-uni-tokens-airdropped-debut/, accessed on October 15, 2022.
[19] Source: https://www.comp.xyz/t/distribution-of-comp-token-to-early-users-pre-comp-distribution-period/, accessed on October 15, 2022.
[20] Source: https://www.comp.xyz/t/should-compound-retroactively-airdrop-tokens-to-early-users/595, accessed on October 15, 2022.

**Figure 5: User acquisition timeline of Compound.**
Panel A plots the number of new unique addresses that deposit tokens into or take loans out of cToken contracts in each day. Of 356,800 addresses, there are about 195,200 addresses that made exactly one stablecoin deposit of $3 or less, with 183,000 addresses that made exactly $3 deposit. The addresses are classified as micro depositors and are excluded from the address-level analysis. There are about 161,500 depositors and 22,300 borrowers in Compound over the sample period. Figure B plots the number of users on a cumulative basis.

Panel A: New depositors and borrowers

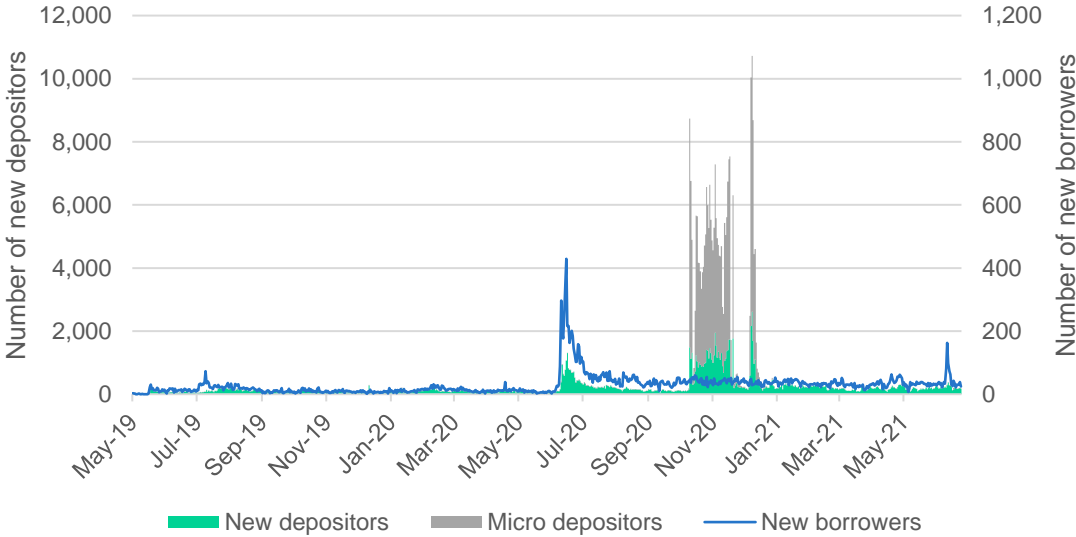

Panel B: Cumulative depositors and borrowers

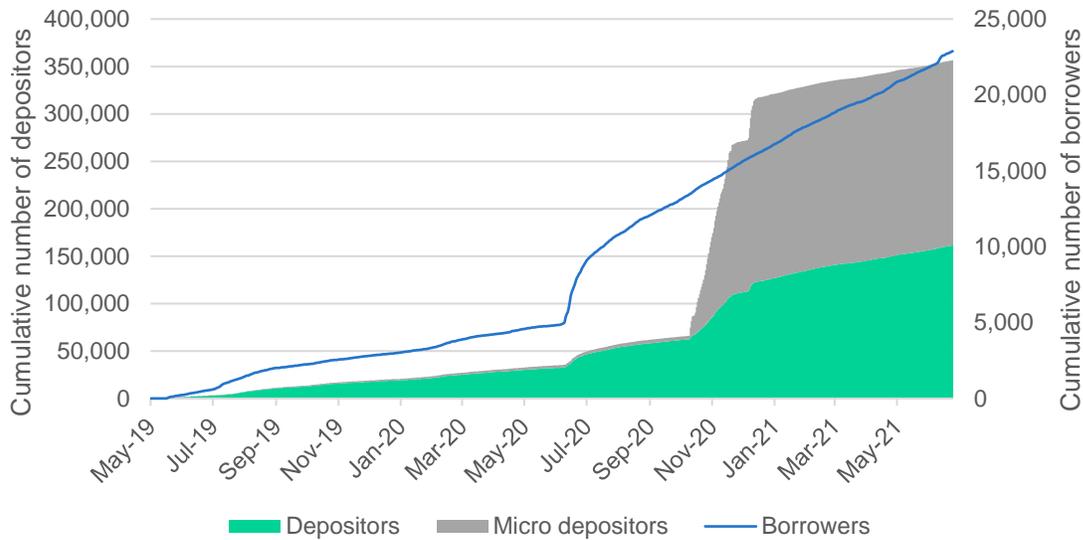

Because of special circumstances surrounding the micro addresses and the small dollar contribution, we exclude them from our analyses. Table 2 presents the summary statistics of the remaining 161,569 addresses, which are classified into 7 categories described earlier. Small addresses account for 99.7% of all addresses but only 16% of dollar value deposited, while the 256 large addresses (0.2%) account for 30.5% of dollar value deposited. Smart contract addresses that are part of DeFi protocols tend to make larger deposits, but there are 109 unidentified contracts that account for most deposits and have the highest average deposits per address. Owners of these contracts do not voluntarily disclose their identity, but as posited in the Appendix, using smart contracts suggest a higher level of sophistication than simple addresses and the average transaction size suggests that they may belong to large investors such as crypto hedge funds. The top 100 addresses account for 75% of all deposits, which is more concentrated than the traditional financial system. For context, Juelsrud (2021) finds that the top 5% of depositors in Norway in 2018 accounted for 53% of all deposits, while OECD data shows that the top 5% of US households owned 68% of total wealth in 2018.[21]

For borrowers, there are 22,289 unique addresses with 52,770 closed loans, which we define as loans that are completely repaid in a complete borrow-repayment cycle, allowing us to calculate loan duration. Most of the dollar value comes from the 217 large addresses, followed by the 51 unidentified contracts. 89% of the loans are drawn in stablecoins, and the top 100 addresses account for 78% of all loans. The average loan duration is 31 days, and longer for stablecoins (33.8 days) than cryptocurrencies (23.2 days). Large addresses tend to borrow for shorter duration than small addresses, and asset management protocols have the highest average loan value and longest duration. This is because the only two asset management contracts in the sample are ETH and Index Coop BTC 2x Flexible Leverage Index, which operate similarly to exchange-traded funds (ETFs) and use USDC loans for leverage.

---

[21] Source: OECD Wealth Distribution Database (WDD), https://stats.oecd.org/Index.aspx?DataSetCode=WEALTH, accessed on October 15, 2022.

Yield aggregators have the second highest average loan value but shortest durations. Larger loan size benefits from economies of scale for gas cost, reducing the friction from fixed adjustment costs and providing greater flexibility in investment strategies. Their short loan duration may result from yield farming rewards that tend to become less lucrative as pool size increases. An analysis in mid-2021 by Nansen, a DeFi analytics service, finds that the 42% of users who entered a liquidity pool on the first day of its launch exited within the first 24 hours, and by the third day, 70% would have withdrawn from the pool.[22] For yield aggregators that are built to automate these strategies, their demand for leverage is likely to be more transient, and this reason also likely applies to large addresses.

The summary statistics for unidentified contracts are closer to yield aggregators than large addresses. For the remaining 11% of the loans that are in cryptocurrencies, their durations are shorter. Overall, the short loan durations make it more likely that Compound loans are used for leveraged investment strategies rather than traditional purposes such as financing.

**Table 2: Summary statistics of Compound users.**
Panel A reports the summary statistics of cToken deposits by address type. Address type classification methodology is outlined in the Appendix. While there are more than 356,800 unique addresses that made deposits, there are about 195,200 addresses that made exactly one stablecoin deposit of 3 USD or less, with 183,000 addresses that made exactly f3 USD deposit identified in Figure 5. These addresses are excluded from our analysis. Total deposits by address type in USD million, average deposits per address in USD million and median deposits in USD are reported. For each address type, the share of addresses that only made stablecoin deposits and addresses that have at least once deposited ETH or WBTC – the two most popular cryptocurrencies – are reported. Panel B reports the number of unique addresses that borrowed via cToken contracts, the number of closed loans (loans that are completely repaid) by token type, and the dollar value of loans in USD million. A closed loan is defined by a complete borrow-repayment cycle for each address; as such, a closed loan can include more than one drawdown and repayment. The average dollar value of loan in USD million and average duration of loan in days are also reported.

Panel A: Depositors.

---

[22] Source: https://www.nansen.ai/research/all-hail-masterchef-analysing-yield-farming-activity, accessed on October 15, 2022.

|  | Total deposits (USD m) | Number of unique addresses | Av. deposits per address (USD m) | Standard Deviation (USD m) | Median value of deposits (USD) |
|---|---|---|---|---|---|
| Large address | 50,725.9 | 256 | 198.1 | 467.3 | 46,700,000 |
| Small address | 26,712.8 | 161,103 | 0.2 | 2.2 | 92 |
| Yield aggregator | 13,018.8 | 41 | 317.5 | 737.7 | 30,100,000 |
| On-ramp | 5,852.2 | 32 | 182.9 | 771.5 | 11,600,000 |
| Decentralized exch. | 3,305.3 | 14 | 236.1 | 476.4 | 36,800,000 |
| Asset management | 721.1 | 14 | 51.5 | 139.5 | 4,618,128 |
| Unidentified contract | 66,226.5 | 109 | 607.6 | 3,922.1 | 16,400,000 |
| All types | 166,562.5 | 161,569 | | | |

|  | Share of deposits (dollar) | Share of addresses (number) | Deposited stablecoins only | Deposited ETH | Deposited WBTC |
|---|---|---|---|---|---|
| Large address | 30.5% | 0.2% | 11% | 75% | 49% |
| Small address | 16.0% | 99.7% | 40% | 49% | 4% |
| Yield aggregator | 7.8% | 0.0% | 61% | 24% | 15% |
| On-ramp | 3.5% | 0.0% | 31% | 66% | 25% |
| Decentralized exch. | 2.0% | 0.0% | 50% | 50% | 43% |
| Asset management | 0.4% | 0.0% | 29% | 14% | 14% |
| Unidentified contract | 39.8% | 0.1% | 25% | 60% | 35% |

Panel B: Borrowers.

|  | Number of unique addresses | Stablecoin loans (num) | Crypto loans (num) | All loans (num) | Stablecoin loans (USD m) | Crypto loans (USD m) | All loans (USD m) |
|---|---|---|---|---|---|---|---|
| Large address | 217 | 1,441 | 614 | 2,055 | 28,659.5 | 3,762.6 | 32,422.1 |
| Small address | 21,986 | 32,220 | 12,230 | 44,450 | 8,164.3 | 1,912.9 | 10,077.2 |
| Yield aggregator | 8 | 55 | 60 | 115 | 4,224.7 | 145.9 | 4,370.6 |
| On-ramp | 24 | 4,062 | 973 | 5,035 | 2,026.6 | 425.6 | 2,452.2 |
| Decentralized exch. | 1 | 7 |  | 7 | 3.8 |  | 3.8 |
| Asset management | 2 | 2 |  | 2 | 256.6 |  | 256.6 |
| Uniden. contract | 51 | 796 | 310 | 1,106 | 11,345.3 | 514.8 | 11,860.0 |
| All types | 22,289 | 38,583 | 14,187 | 52,770 | 54,680.7 | 6,761.9 | 61,442.6 |

|  | Average value of loan (USD m) | | | Average loan duration (days) | | |
|---|---|---|---|---|---|---|
|  | Stable | Crypto | All | Stable | Crypto | All |
| Large address | 19.89 | 6.13 | 15.78 | 23.6 | 15.2 | 21.1 |
| Small address | 0.25 | 0.16 | 0.23 | 40.6 | 26.6 | 36.7 |
| Yield aggregator | 76.81 | 2.43 | 38.01 | 2.4 | 2.9 | 2.7 |
| On-ramp | 0.50 | 0.44 | 0.49 | 0.6 | 0.3 | 0.6 |
| Decentralized exch. | 0.54 |  | 0.54 | 0.0 |  | 0.0 |
| Asset management | 128.32 |  | 128.32 | 81.6 |  | 81.6 |
| Uniden. contract | 14.25 | 1.66 | 10.72 | 9.0 | 6.5 | 8.3 |
| All types | 1.42 | 0.48 | 1.16 | 33.8 | 23.2 | 31.0 |

## 3.2 Hypothesis Development and Empirical Methodology

As discussed in Section 2.2, Compound follows the intermediation banking model, connecting depositors and lenders. Our first and most basic hypothesis is that users respond to interest rates, which are algorithmically determined by the interest rate model and automatically enforced by smart contracts.

*H1: Demand for deposits and loans in Compound is influenced by supply and borrow interest rates.*

In an analogue to traditional money market funds, Compound may be used to earn income, which is an improvement on DeFi alternative at that time of simply holding tokens on the blockchain. For daily net deposits (deposits minus withdrawals, which represent net inflows), the variable could be negative, representing net withdrawals. Thus, the dependent variable is measured in USD million, as reported in Table 1 Panel A. The main determinant is the lagged supply rate (SupplyRate), earlier described in Equation 2. Rate variables determined by Compound's interest rate model need to be lagged because they are in turn determined by the net flows into and out of the cToken contracts, leading to reverse causality.

We control for size of the cToken contract with log of deposits in USD million and the three dimensions of market conditions: past 1-day return (price change), past 7-day return, and past 30-day volatility of the crypto market, represented by the vector $X_t$. The correlations between ETH and other tokens are between 0.73 and 0.92, while the correlation between ETH and BTC is 0.80. Because the dominant deposited tokens are stablecoins, ETH and WBTC, and the protocol is built on the Ethereum blockchain, we use control variables computed based only on ETH to avoid multicollinearity.

Because Compound started distributing COMP on June 15, 2020, this reward may attract inflows into the protocol. Two weeks after the introduction, Compound team was "surprised by how powerful the impact of the distribution was on incentives, and so was the community",[23] which is corroborated by a sharp rise in borrowing activities as illustrated earlier in Figure 3. To analyze how users' sensitivity to supply and borrow rates are affected, we include an interaction term to allow the coefficients on lagged supply rate to change after COMP distribution and include lagged COMP distribution reward for suppliers (SupplyReward) computed as annual percentage rate (APY) in the regression. For periods prior to COMP distribution, SupplyReward takes value of zero. If borrowers are attracted by COMP distribution, the coefficient on SupplyReward should be positive.

---

[23] Source: https://www.coindesk.com/tech/2020/06/30/compound-changes-comp-distribution-rules-following-yield-farming-frenzy/, accessed on October 15, 2022.

*H2: Demand for deposits and loans in Compound is influenced by COMP reward distribution.*

The daily net deposit regression follows Equation 4, where the subscript $i$ for cToken $i$ is suppressed for brevity. Standards errors are estimated using the Newey-West procedure with one-day lag to account for potential serial correlation in the data.

$$NetDeposit_t = \alpha + \beta_0 Post_{t-1} + \beta_1 SupplyRate_{t-1} + \beta_2 Post_{t-1} \times SupplyRate_{t-1} \quad (4)$$
$$+ \beta_3 SupplyReward_{t-1} + \beta_4 \log(PoolSize_t) + \gamma X_t + \varepsilon_t$$

Stablecoins offer a more stable store of wealth (in nominal dollar) during uncertain times, which can lead to increased demand similar to flight to safety (Baele et al. 2020). We refer to this as the safety demand hypothesis. Under the safety demand hypothesis, we should observe a positive relationship between net deposits and past 30-day volatility.

*H3: During periods of high volatility, demand for stablecoin deposits increases.*

We repeat similar analyses for token loans, replacing the supply rates with borrow rates (BorrowRate, earlier described in Equation 1, and BorrowReward), and the dependent variable with log of 1 plus loan amount in USD million, since all values are non-negative. The loan regression follows Equation 5. For both net deposits and loans, rather than aggregating the tokens into 2 broad categories or rely on panel data estimation technique, we separate each token and analyze the data as time series because their underlying properties and roles in the DeFi ecosystem are different, so the sensitivities and determinants can be different.

$$\log(1 + Loan_t) = \alpha + \beta_0 Post_{t-1} \quad (5)$$
$$+ \beta_1 BorrowRate_{t-1} + \beta_2 Post_{t-1} \times BorrowRate_{t-1}$$
$$+ \beta_3 BorrowReward_{t-1} + \beta_4 \log(PoolSize_t) + \gamma X_t + \varepsilon_t$$

The loan demand in DeFi could be driven by (1) idiosyncratic demand for liquidity without necessitating a sale that triggers capital gains tax, (2) demand for leverage for long positions in cryptocurrencies, (3) demand for shorting via repurchase agreement, or (4) demand for yield farming. For channel (1) and (2), loan demand would be higher in bullish market conditions and loans would more likely be drawn in stablecoins. For channel (3), loans would more likely be drawn in non-stablecoins since token depreciation would make borrowing cost lower, and demand would be higher in bearish market conditions. For channel (4), loan demand would be responsive to borrow reward rate. Without detailed financial positions of each borrower, it is difficult to discern the demand channels, except for yield farming where we can gain more accurate insights directly from Compound. Because Compound itself offers internal yield farming opportunities, we can identify these borrowers from those who redeposit borrowed tokens to Compound to recursively earn COMP, which is possible with blockchain transaction data.

For loan demand analysis, ETH's volatility has a different interpretation. Most cryptocurrencies exhibit positive co-movements, so stablecoin loans are likely to face the greatest asset-liability price mismatch risk. While we are unable to observe the token collaterals that back each loan, if one assumes that borrowers are more likely to rely on cryptocurrencies as collateral more than stablecoins, then periods where ETH's price is more volatile would place the borrower at greater risk of liquidation, and thus would be less likely to take out a stablecoin loan. Consequently, we expect to see a negative coefficient on ETH's past 30-day volatility.

*H4: During periods of high volatility, demand for stablecoin loans decreases.*

To further investigate yield farming, we refer to the long-short, liquidation-free strategy described in Section 2.4, where borrowers would redeposit their borrowed tokens into the cToken contract. We use address-level data to identify redeposits and infer the demand for leveraged yield farming. In addition to univariate analyses, we fit a logistic regression of the likelihood of an address redepositing borrowed tokens on similar variables, but also adding the size of the token loan drawn in USD and fixed effects for address type. The regressions are estimated for the top five tokens in dollar value: ETH, WBTC, DAI, USDC, and USDT, and standard errors are clustered by addresses.

## 4. Results
### 4.1 Determinants of Net Deposits and Loan Demand

In Figure 5, we already saw that COMP distribution may affect both the demand for deposits and loans as new borrower and depositor addresses were created after June 15. To investigate the hypotheses more systematically, we turn to the multivariate regressions. We display the result of each cToken contract in a separate column beginning with the eight cryptocurrencies followed by the four stablecoins.

Table 3 reports the determinants of net deposits. For most tokens, the adjusted R-squared values are close to zero or negative, and there is no clear common pattern across all tokens, corroborating the view that tokens are heterogenous assets. Unlike traditional financial services that restrict the types of assets that can be deposited or borrowed, DeFi allows for much greater flexibility. This lack of systematic pattern is true even for stablecoins, which despite being broadly classifiable as tokens intended to track the price of a USD, are created in different ways, by different developers, and can serve different roles in the DeFi ecosystem.

For example, net deposits are only positively related to supply rates for WBTC, BAT, and USDT, while for REP, the relationship is negative. The introduction of COMP reward also seems to have little effect, with only BAT and COMP reporting positive but statistically weak relationship to supply rewards. This finding is consistent with the pattern observed in Figure 5. There is also no clear systematic relationship between market conditions and net deposits, suggesting that the safety demand deposit cannot be substantiated.[24]

We conduct a further investigation by aggregating net deposits by address type, focusing on the top five tokens identified in Figure 2 (ETH, WBTC, DAI, USDC, and USDT). The results are reported in Table A1 of the Appendix. We find that some patterns emerge: for example, net deposits for small addresses tend to respond positively to supply rates more (WBTC, USDC, and USDT), which could be interested as saving demand, while yield aggregator and asset management protocols' ETH net deposit is positively related to market conditions (7-day return, 30-day volatility), which may be a result of fund inflows during bull market. However, the adjusted R-squared values remain low, and no common pattern emerges. In conclusion, the aggregate behavior of depositors does not seem to exhibit any clear systematic relationship.

---

[24] In a previous version of this paper, two clear patterns that emerges are that (1) net deposit in cDAI contract is positively and strongly related to net increase in DAI stablecoins issued by MakerDAO, resulting in very high adjusted R-squared, and (2) net deposit in cBAT contract is negatively related to net increase in DAI, which could be due to competition from MakerDAO which began accepting BAT as collateral to mint DAI in late 2019. We remove these findings from this version of the paper because of the tokens' specificity, but the results highlight the potential rivalry between DeFi protocols for users' liquidity, and the different roles that each token may have in the DeFi ecosystem.

While there are many idiosyncratic reasons why one might deposit, the circumstances surrounding loan demands are more defined. However, the regression results also seem to be puzzling. Table 4 reports the determinants of loan demands. Contrary to the typical interpretation of interest rate as loan price, for 6 out of the 12 tokens, the relationship with lagged borrow rate is positive, with only TUSD reporting negative relationship. To make sense of this result, we must also consider the relationship between loan demands and borrow reward in conjunction. The coefficients on borrow reward rates are unanimously positive, with 8 out of the 10 tokens that receive COMP distribution showing statistically significant coefficients.

Recall that interest rates automatically increase when utilization rate is low or loan demand is high, unlike traditional lending where interest rates do not algorithmically and immediately adjust to changes in the lender's liquidity position. Under this interpretation, borrowers are willing to pay high interest rate as the supply of loanable funds depletes. As described in Section 2.4, it is possible to execute an arbitrage strategy when COMP reward is taken into account. In fact, the net borrowing rate needs not be negative, as users receive both interest income and reward from depositing initial collateral and redeposited tokens into Compound.

This practice is referred to as "leveraged yield farming",[25] and as soon as one week after then introduction of COMP reward in June 2020, an article remarked that this potential for recursive interaction "triggered a gold rush of yield arbitrage, sending its assets under management and price to new heights".[26] This would also explain why loan demand in TUSD, a stablecoin which not eligible for COMP reward, is negatively related to borrow rate as one might expect. And for the case of leveraged yield farming outside Compound, this also holds true: as long as the borrowed funds can be deployed in more profitable strategies, the borrow rates do not matter. And as users borrow more, borrow rates increase more, which can result in the positive correlation observed in the data. It is the loan demand that drives up the interest rate, and the increased interest rate is not sufficient to dampen loan demand.

---

[25] A description of leveraged yield farming strategy in Compound could be found in https://defiprime.com/defi-yield-farming, accessed on November 11, 2022.
[26] Source: https://cointelegraph.com/news/compound-reward-farming-results-in-six-fold-increase-of-lending-activity, accessed on November 11, 2022.

As discussed in hypothesis 4, stablecoin loans face greater risk of liquidation during volatile times. We see the negative relationship between loan demands and past 30-day standard deviation of ETH for USDC, USDT, and TUSD. The adjusted R-square values for the stablecoin loans are also higher than cryptocurrency loans.

In Table A2 in the Appendix, we separate addresses by type and find that the positive relationship with borrow rates and borrow rewards still hold, but mostly for large addresses and small addresses. Protocols and unidentified contracts exhibit positive relationship to borrow rates, but the relationship with borrow rewards are unclear. Stablecoin loans by large addresses and small addresses are also less likely during volatile markets, while the behaviors of protocols are more diverse. Under the liquidation risk hypothesis, if we assume that addresses are more likely to be manually operated while protocols are automated, this finding is consistent with the view that addresses are more averse to liquidation risk and are therefore less likely to take stablecoin loans during volatile times. It may come as surprise that protocols do not respond very strongly to borrow reward rates (in fact, negatively for USDC). It is possible that there are more profitable rewards to be obtained in other protocols or complex deployable strategies and thus their objectives may be to use Compound simply to gain leverage, not for its reward.

Thus, the positive relationship between borrow rate and loan demand is likely explained by leveraged yield farming, both within Compound and in other protocols in the ecosystem. Since deposits are an important part of such strategies, this could also explain why much of the activities are not explainable by typical savers' behavior. In Table A1 Panel E, we analyze the net deposit behavior of addresses that also borrow and find similar a lack of systemic relationship. In other words, DeFi yield farming is likely the main use case and driver of growth for Compound. In the next section, we further investigate leveraged yield farming within Compound further from address-level activities.

**Table 3: Determinants of net deposits.**
This table reports the result from the regressions of daily net deposits (deposits minus withdrawals) in USD million between May 2019 and June 2021 for the 12 accepted tokens. In column 1 to 8, the results for cryptocurrencies are reported, followed by stablecoins from column 8 to 12. Explanatory variables include lagged supply rate, its interaction with an indicator for periods post COMP distribution, lagged supply COMP reward rate, log dollar value of deposits in USD million, 1-day ETH return, 7-day ETH return, and 30-day ETH volatility (measured in percentage point). Standard errors are computed using the Newey-West procedure with one-day lag and reported in parenthesis. Stars correspond to statistical significance level, with *, ** and *** representing 10%, 5% and 1% respectively.

| VARIABLES | (1) cETH | (2) cWBTC | (3) cBAT | (4) cCOMP | (5) cLINK | (6) cREP | (7) cUNI | (8) cZRX | (9) cDAI | (10) cUSDC | (11) cUSDT | (12) cTUSD |
|---|---|---|---|---|---|---|---|---|---|---|---|---|
| Post COMP distribution | -6.63 | 5.62* | 0.257 | | | -0.401*** | | 0.051 | -34.31 | -18.46 | 2.95 | |
|  | (5.90) | (2.90) | (0.440) | | | (0.130) | | (0.380) | (28.28) | (14.42) | (2.69) | |
| Lagged supply rate | -14.35 | 1.80*** | 0.685*** | 0.024 | -1.98 | -1.96*** | -0.110 | -0.021 | -7.14 | -2.86 | 1.88** | 5.00 |
|  | (19.36) | (0.260) | (0.180) | (0.020) | (1.33) | (0.360) | (0.690) | (0.120) | (5.43) | (2.21) | (0.880) | (8.85) |
| Post * lagged supply rate | 63.72 | 4.74 | -0.780*** | | | | | -0.185 | 11.14* | 4.80 | -0.978 | |
|  | (59.98) | (6.12) | (0.230) | | | | | (0.200) | (6.29) | (3.45) | (0.910) | |
| Lagged supply reward | 1.11 | -3.66 | 0.168* | 0.312* | 1.00 | | 1.22 | 0.124 | 2.10 | 7.55 | 0.017 | |
|  | (13.21) | (4.08) | (0.10) | (0.18) | (0.66) | | (1.46) | (0.080) | (3.29) | (5.84) | (0.410) | |
| log(pool size) | -0.002 | -0.00 | -0.006 | -0.006* | -0.064 | -0.020** | -0.002 | -0.001 | -0.005 | -0.004 | -0.009 | -0.408 |
|  | (0.000) | (0.000) | (0.010) | (0.000) | (0.060) | (0.010) | (0.010) | (0.000) | (0.010) | (0.010) | (0.010) | (0.270) |
| ETH return (1d) | 20.76 | -64.07 | -3.94 | -6.89 | 2.18 | -0.646 | -20.69 | -0.792 | 18.97 | -40.55 | -33.13* | -25.71 |
|  | (52.57) | (39.26) | (2.58) | (6.31) | (16.61) | (0.480) | (16.18) | (1.00) | (133.04) | (144.62) | (17.55) | (34.90) |
| ETH return (7d) | 5.28 | 28.40 | 0.444 | -2.50 | 3.98 | -0.439** | 2.40 | 0.661 | 87.13 | -7.18 | 8.32 | -4.02 |
|  | (25.61) | (19.84) | (1.15) | (2.07) | (14.09) | (0.170) | (3.19) | (0.690) | (64.69) | (72.60) | (7.77) | (7.19) |
| ETH SD (30d) | -27.53 | -42.44 | -0.408 | 3.20 | 65.29 | -1.68 | 21.79 | -0.300 | -15.07 | 31.84 | 86.10 | -279.81 |
|  | (225.17) | (152.05) | (6.56) | (14.35) | (44.16) | (1.12) | (34.25) | (3.48) | (392.15) | (501.76) | (59.86) | (415.90) |
| Constant | 3.78 | -1.21 | 0.06 | 0.145 | 0.087 | 0.506*** | -1.205 | 0.231 | 21.91 | 5.06 | -8.94*** | 51.60 |
|  | (9.12) | (5.27) | (0.490) | (0.530) | (4.46) | (0.170) | (1.74) | (0.280) | (19.30) | (19.24) | (2.88) | (40.23) |
| Observations | 426 | 351 | 426 | 256 | 39 | 426 | 270 | 426 | 426 | 426 | 425 | 40 |
| Adj R-squared | -0.007 | -0.010 | 0.305 | 0.053 | -0.070 | 0.041 | -0.006 | -0.010 | 0.008 | -0.008 | 0.080 | 0.217 |

**Table 4: Determinants of loans drawn.**

This table reports the result from the regressions of log cToken loan between May 2019 and June 2021. In column 1 to 8, the results for cryptocurrencies are reported, followed by stablecoins from column 8 to 12. Explanatory variables include lagged borrow rate, its interaction with an indicator for periods post COMP distribution, lagged borrow COMP reward rate, log dollar value of deposits in USD million, 1-day ETH return, 7-day ETH return, and 30-day ETH volatility (measured in percentage point). Standard errors are computed using the Newey-West procedure with one-day lag and reported in parenthesis. Stars correspond to statistical significance level, with *, ** and *** representing 10%, 5% and 1% respectively.

| VARIABLES | (1) cETH | (2) cWBTC | (3) cBAT | (4) cCOMP | (5) cLINK | (6) cREP | (7) cUNI | (8) cZRX | (9) cDAI | (10) cUSDC | (11) cUSDT | (12) cTUSD |
|---|---|---|---|---|---|---|---|---|---|---|---|---|
| Post COMP distribution | 3.36** | -2.57*** | 0.504 | | | -1.84 | | -2.21 | -2.39*** | 0.875 | 3.06*** | |
|  | (1.57) | (0.910) | (0.650) | | | (2.34) | | (3.13) | (0.46) | (0.580) | (0.950) | |
| Lagged borrow rate | 2.66*** | 0.131* | 0.276*** | 0.070* | -0.185 | 0.858 | 0.085 | 0.342*** | 0.064 | 0.093 | 0.285*** | -3.29*** |
|  | (0.490) | (0.070) | (0.030) | (0.040) | (0.260) | (0.880) | (0.080) | (0.120) | (0.060) | (0.100) | (0.080) | (0.780) |
| Post * lagged borrow rate | -1.29** | 0.147 | -0.050 | | | -0.750 | | -0.170 | 0.002 | -0.104 | -0.252*** | |
|  | (0.610) | (0.14) | (0.060) | | | (0.880) | | (0.230) | (0.060) | (0.100) | (0.080) | |
| Lagged borrow reward | 0.061* | 0.089** | 0.019*** | 0.006 | 0.203** | | 0.440*** | 0.187* | 0.110*** | 0.032 | 0.096** | |
|  | (0.030) | (0.04) | (0.000) | (0.18) | (0.100) | | (0.090) | (0.110) | (0.030) | (0.030) | (0.040) | |
| log(pool size) | 0.756*** | 0.524*** | 0.854*** | -3.01*** | -0.533 | 0.491*** | -0.767* | 0.297 | 1.23*** | 1.19*** | 1.10*** | 0.497*** |
|  | (0.080) | (0.12) | (0.230) | (0.74) | (1.87) | (0.150) | (0.420) | (0.370) | (0.08) | (0.080) | (0.120) | (0.110) |
| ETH return (1d) | 1.14 | -3.42* | -2.52 | 1.33 | -8.80 | -0.426 | 6.22 | -1.12 | 0.542 | -0.074 | -1.80 | 2.38 |
|  | (1.43) | (2.07) | (4.91) | (5.51) | (6.02) | (1.84) | (4.87) | (3.81) | (1.44) | (1.08) | (1.36) | (3.46) |
| ETH return (7d) | -0.310 | 1.50 | -3.56 | 6.27** | 9.87*** | -3.33*** | -3.80 | -2.62 | 0.302 | 0.484 | 1.71** | -0.195 |
|  | (0.670) | (1.07) | (2.22) | (2.65) | (3.01) | (0.920) | (2.35) | (2.21) | (0.670) | (0.530) | (0.700) | (1.84) |
| ETH SD (30d) | 4.61 | 30.25*** | -21.92 | 14.96 | 15.36 | -8.55* | -23.66 | -10.90 | -0.540 | -16.31*** | -21.61*** | -110.11* |
|  | (5.45) | (9.72) | (16.31) | (19.87) | (38.64) | (4.50) | (18.09) | (14.85) | (4.64) | (3.77) | (5.28) | (62.09) |
| Constant | 1.40 | 9.04*** | 4.75*** | 16.68*** | 12.95 | 1.53 | 11.65*** | 1.76 | 9.43*** | 8.76*** | 7.48*** | 22.75*** |
|  | (1.20) | (0.80) | (0.78) | (1.69) | (8.88) | (2.33) | (1.95) | (1.22) | (0.47) | (0.61) | (0.88) | (6.14) |
| Observations | 426 | 351 | 426 | 256 | 39 | 426 | 270 | 426 | 426 | 426 | 425 | 40 |
| Adj R-squared | 0.552 | 0.209 | 0.255 | 0.395 | 0.194 | 0.373 | 0.065 | 0.101 | 0.583 | 0.727 | 0.657 | 0.688 |

### 4.2 Recursive Interactions

In subsequent analyses, we use transaction data to investigate users' interactions at a more granular level. As blockchain data is unorganized and understanding the full financial positions of each address requires manually parsing all on-chain transactions, we limit the scope of our analysis to interactions with Compound only. As discussed earlier, a leveraged yield farmer within Compound would redeposit their borrowed tokens, so we can identify addresses that exhibit this behavior to gain a better understanding of who they are.

To classify such addresses, we reaggregate the data by address loan day, which is defined by aggregating all loans drawn from a token by an address in each day. Since the loans are not required to be closed like in Table 2, and each closed loan may involve multiple draw spread over multiple days, there are more loan day observations (98,717) than closed loans (52,770). The summary statistics of loan days and their redeposit rates are reported in Table 5.

We analyze the number of the loans, not the dollar value. On average, 11.4% of cToken loans are redeposited within one day, and 10.5% within the same day, with ETH (20.6%) and BAT (28.5%) as the most redeposited tokens followed by COMP (18.5%) and UNI (15.9%). Of the four stablecoins, DAI is the most redeposited (14.7%), followed by USDC (7.4%). Across address types, yield aggregators and unidentified contracts are most likely to redeposit, consistent with their objectives as outlined in the Appendix. This evidence may seem contradictory to the result from loan demand analysis, but it is important to note that this analysis is conducted on the frequencies, not the dollar proportion. Excluding on-ramps and decentralized exchanges, small addresses are the least likely to redeposit, which may be due to their lower level of sophistication or relatively smaller position sizes that makes the benefits not worth the gas cost.

Next, we investigate the determinants of loan redeposits more systematically by a logistic regression whose dependent variable is the redeposit dummy variable. The dummy takes value of one if the address redeposits the borrowed taken within one day. We pool all types of address and control for the cross-category differences in redeposit rates observed in Table 5 by address type fixed effects. Because leveraged yield farming strategies earn rewards from both supplying and borrowing, we include the determinants from Equation 4 and 5, as well as log of loan size in USD. Table 6 reports the estimated coefficients of the regressions for the five main tokens.

First of all, it is worth noting that there are far more stablecoin loans than cryptocurrency loans, and their pseudo R-squared values are also larger. The relationships for cryptocurrency loans are statistically weaker, and only the supply rewards, not borrow rewards, are positively related to deposits. From Table 1 Panel C, the borrow rewards for ETH and WBTC are high due to low utilization, and for much of the sample period, the long-short, liquidation-free strategy described in Section 2.4 would offer a risk-free arbitrage.[27] Thus, changes in reward rates would likely have little impact on whether borrowed tokens are more likely to be redeposited.

For stablecoin loans, DAI and USDC redeposits are more responsive to rates, both the borrowing spread and rewards. USDT redeposit, on the other hand, is only positively related to borrow rate. Larger addresses are also more likely to redeposit borrowed tokens, consistent with how the profitability of leveraged yield farming would be attenuated by gas cost which does not vary much by transaction size.

Finally, in Table 7, we analyze the behavior of the addresses that sent instructions to claim COMP rewards, which only amount to 16,968 addresses from the 161,569 users, excluding the micro addresses identified in Figure 5. While most interactions with Compound are eligible for reward, gas cost can make claiming and monetizing COMP uneconomical for small transactions. While there are 873,652 COMP claimed during this period, more than 1.9 million are deposited into cCOMP contract (labelled as internal yield farming in Table 7), and almost 0.42 million sent to other yield aggregator or asset management protocols (labelled as external yield farming). This discrepancy arises because the deposits can be made from COMP that were claimed or bought. Relative to their claimed amount, there are much more internal and external yield farming activities by small address, indicating larger turnover relative to other types.

On a per address basis, unidentified contracts claim almost as much as large addresses, and much more than yield aggregators, which suggests that some contracts may possibly be part of protocols but are not explicitly labeled or identified to users. In Panel B, we restrict the analysis to addresses that redeposit borrowed tokens back into the relevant cToken contract. Addresses that redeposit account for 28.2% of loans drawn from Compound. The number of addresses decline by 85% to 2,551 addresses but they account for 76.3% of COMP claimed. The proportion of addresses that engaged in yield farming also increase from 13% to 19% for internal and 5.7% to 7.8% for external. In sum, evidence from our loan-level and address-level analyses suggest that yield farming is prevalent, but the dollar value is concentrated among less than 1% of users.

---

[27] In fact, this is a strategy employed by Yearn, a yield aggregator protocol. A description of Yearn's GenLevCompV2 strategy in WBTC yVault reads "Supplies and borrows yvWBTC on Compound Finance simultaneously to earn COMP. Earned tokens are harvested, sold for more yvWBTC which is deposited back into the strategy. Flashloans are used to obtain additional yvWBTC from dYdX to boost the APY." This also explains why 97.3% of WBTC is redeposited by yield aggregator addresses in Table 5. Source: https://yearn.watch/vault/0xA696a63cc78DfFa1a63E9E50587C197387FF6C7E/0x619Dde92f9fD8Af679025A2fD7e9ED2269e4c0c8, accessed November 11, 2022.

**Table 5: Redeposited loans.**

Panel A reports the number of loan day by token and address type. A loan day is defined as a day that an address takes out a token loan, which may involve multiple draws within the same day. Because of this, the number of loan days is higher than the number of closed loans reported in Table 2, which are defined as complete borrow-repayment cycle. Panel B reports the share of loan days where the borrower immediately redeposits the borrowed tokens to the cToken contract on the same day. Panel C reports the share of loan days where a redeposit occurs within one day.

Panel A: Distribution of daily loans by token

|  | cETH | cWBTC | cBAT | cCOMP | cLINK | cREP | cUNI | cZRX | cDAI | cUSDC | cUSDT | cTUSD | Total |
|---|---|---|---|---|---|---|---|---|---|---|---|---|---|
| Large address | 348 | 167 | 141 | 45 | 30 | 20 | 72 | 36 | 2,817 | 1,859 | 1,048 | 10 | 6,594 |
| Small address | 8,026 | 2,280 | 2,827 | 418 | 214 | 371 | 765 | 772 | 31,177 | 27,934 | 14,344 | 327 | 89,539 |
| Yield aggregator |  | 37 |  |  |  |  |  |  | 2 | 9 | 2 |  | 50 |
| On-ramp |  |  |  |  |  |  |  |  |  | 180 |  |  | 180 |
| Decentralized exchange |  |  |  |  |  |  |  |  | 3 |  |  |  | 3 |
| Asset management | 32 | 5 | 21 |  |  | 6 | 11 | 15 | 113 | 112 | 38 |  | 357 |
| Unidentified contract | 50 | 117 | 50 | 8 | 8 | 19 | 25 | 32 | 1,045 | 444 | 187 |  | 1,994 |
| All types | 8,456 | 2,606 | 3,039 | 471 | 252 | 416 | 873 | 855 | 35,157 | 30,538 | 15,619 | 337 | 98,717 |

Panel B: Share of daily loans that are redeposited on the same day

|  | cETH | cWBTC | cBAT | cCOMP | cLINK | cREP | cUNI | cZRX | cDAI | cUSDC | cUSDT | cTUSD | Total |
|---|---|---|---|---|---|---|---|---|---|---|---|---|---|
| Large address | 13.5% | 22.2% | 49.6% | 33.3% | 40.0% | 10.0% | 13.9% | 16.7% | 14.3% | 9.8% | 1.2% | 10.0% | 12.1% |
| Small address | 19.1% | 10.5% | 24.6% | 17.0% | 9.8% | 4.3% | 15.3% | 8.4% | 12.5% | 6.0% | 2.5% | 0.3% | 9.7% |
| Yield aggregator |  | 97.3% |  |  |  |  |  |  | 0.0% | 0.0% | 0.0% |  | 72.0% |
| On-ramp |  |  |  |  |  |  |  |  |  | 0.0% |  |  | 0.0% |
| Decentralized exchange |  |  |  |  |  |  |  |  | 100.0% |  |  |  | 100.0% |
| Asset management | 12.5% | 0.0% | 23.8% |  |  | 0.0% | 0.0% | 0.0% | 55.8% | 8.9% | 0.0% |  | 23.0% |
| Unidentified contract | 34.0% | 20.5% | 54.0% | 0.0% | 62.5% | 52.6% | 16.0% | 12.5% | 47.5% | 42.8% | 1.6% |  | 39.1% |
| All types | 18.9% | 12.9% | 26.3% | 18.3% | 15.1% | 6.7% | 15.0% | 8.8% | 13.8% | 6.8% | 2.4% | 0.6% | 10.5% |

Panel C: Share of daily loans that are redeposited within one day

|  | cETH | cWBTC | cBAT | cCOMP | cLINK | cREP | cUNI | cZRX | cDAI | cUSDC | cUSDT | cTUSD | Total |
|---|---|---|---|---|---|---|---|---|---|---|---|---|---|
| Large address | 16.4% | 24.0% | 51.1% | 33.3% | 43.3% | 20.0% | 16.7% | 25.0% | 15.7% | 10.8% | 2.0% | 10.0% | 13.5% |
| Small address | 20.7% | 12.2% | 26.8% | 17.2% | 10.7% | 5.4% | 16.1% | 10.1% | 13.3% | 6.6% | 3.0% | 0.3% | 10.5% |
| Yield aggregator |  | 97.3% |  |  |  |  |  |  | 0.0% | 0.0% | 0.0% |  | 72.0% |
| On-ramp |  |  |  |  |  |  |  |  |  | 0.0% |  |  | 0.0% |
| Decentralized exchange |  |  |  |  |  |  |  |  | 100.0% |  |  |  | 100.0% |
| Asset management | 12.5% | 0.0% | 38.1% |  |  | 0.0% | 0.0% | 0.0% | 57.5% | 11.6% | 0.0% |  | 25.2% |
| Unidentified contract | 34.0% | 21.4% | 58.0% | 0.0% | 62.5% | 63.2% | 16.0% | 15.6% | 48.9% | 44.1% | 3.2% |  | 40.6% |
| All types | 20.6% | 14.5% | 28.5% | 18.5% | 16.3% | 8.7% | 15.9% | 10.8% | 14.7% | 7.4% | 2.9% | 0.6% | 11.4% |

**Table 6: Determinants of redeposited loans.**
This table reports the result from the logistic regression of redeposit dummy variable (which takes value of 1 when the address redeposits borrowed tokens within 1 day) on log of loan size in USD, lagged supply rate, lagged borrow rate, their interactions with an indicator for periods post COMP distribution, lagged supply COMP reward rate, lagged borrow COMP reward rate, 1-day ETH return, 7-day ETH return, 30-day ETH volatility (measured in percentage point), and address type fixed effects. Only the 5 main tokens (ETH, WBTC, DAI, USDC, and USDT) are analyzed. Standard errors are clustered by address and reported in parenthesis. Stars correspond to statistical significance level, with *, ** and *** representing 10%, 5% and 1% respectively.

|  | (1) ETH | (2) WBTC | (3) DAI | (4) USDC | (5) USDT |
|---|---|---|---|---|---|
| Log(loan size in USD) | -0.024* | 0.066** | 0.224*** | 0.235*** | 0.066** |
|  | (0.010) | (0.030) | (0.020) | (0.020) | (0.030) |
| Post COMP distribution | 2.36 | -1.14 | -0.946*** | -1.11** | -0.927 |
|  | (2.84) | (2.21) | (0.15) | (0.440) | (0.620) |
| Lagged supply rate | -3.71 | 0.392 | -0.338*** | -0.008 | -0.084 |
|  | (4.66) | (0.91) | (0.10) | (0.18) | (0.06) |
| Lagged borrow rate | 2.01 | -0.307 | 0.296*** | -0.065 | 0.204*** |
|  | (1.27) | (0.590) | (0.090) | (0.14) | (0.07) |
| Post * lagged supply rate | -6.378 | -2.601 | 1.809*** | -0.502** | 0.082 |
|  | (5.47) | (1.69) | (0.30) | (0.24) | (0.24) |
| Post * lagged borrow rate | -0.471 | 0.503 | -1.26*** | 0.449** | -0.167 |
|  | (1.36) | (0.62) | (0.22) | (0.20) | (0.18) |
| Lagged supply reward | 1.48* | 1.05* | -3.77*** | 0.217*** | -0.087 |
|  | (0.850) | (0.610) | (0.390) | (0.080) | (0.200) |
| Lagged borrow reward | -0.076 | -0.032 | 3.01*** | -0.104 | 0.139 |
|  | (0.050) | (0.040) | (0.300) | (0.060) | (0.130) |
| ETH return (1d) | -0.174 | 0.995 | 0.105 | -0.273 | -3.58** |
|  | (0.500) | (1.09) | (0.320) | (0.350) | (1.50) |
| ETH return (7d) | 0.186 | -0.584 | -0.606*** | -0.422** | -0.987 |
|  | (0.240) | (0.550) | (0.190) | (0.200) | (0.760) |
| ETH SD (30d) | -0.848 | 9.85 | -15.15*** | -0.987 | -14.77 |
|  | (2.09) | (6.07) | (2.05) | (1.67) | (10.19) |
| Address type fixed effects | Yes | Yes | Yes | Yes | Yes |
| Observations | 8,514 | 1,874 | 35,378 | 30,601 | 15,704 |
| Pseudo R-squared | 0.030 | 0.048 | 0.116 | 0.084 | 0.212 |

**Table 7: COMP activities.**
Panel A reports the number of addresses that claimed COMP rewards by address type. COMP claimed (in units), sent for internal yield farm by depositing into cCOMP contract, and sent for external yield farming to yield aggregator or asset management protocols are reported, first as aggregate by address type and then again as average per address. Finally, the proportion of addresses that yield farm their COMP tokens are reported. Panel B repeats the summary statistics for addresses that redeposit their borrowed tokens back into cToken contracts.

Panel A: All addresses

|  | Number of addresses | Total COMP activities | | | Average COMP activities | | | % yield farming | |
| --- | --- | --- | --- | --- | --- | --- | --- | --- | --- |
|  |  | Claimed | Internal | External | Claimed | Internal | External | Internal | External |
| Large address | 192 | 475,048 | 842,865 | 238,255 | 2,474.2 | 4,389.9 | 1,240.9 | 34.4% | 15.1% |
| Small address | 16,661 | 242,605 | 1,016,698 | 178,836 | 14.6 | 61.0 | 10.7 | 12.7% | 5.5% |
| Yield aggregator | 23 | 10,665 | 48,754 | 131 | 463.7 | 2,119.7 | 5.7 | 8.7% | 17.4% |
| On-ramp | 8 | 1,995 | 6,351 | 2,434 | 249.4 | 793.8 | 304.3 | 25.0% | 37.5% |
| Decentralized exchange | 1 | 547 | 0 | 0 | 546.9 | 0.0 | 0.0 | 0.0% | 0.0% |
| Asset management | 24 | 8,625 | 1,278 | 0 | 359.4 | 53.3 | 0.0 | 16.7% | 0.0% |
| Unidentified contract | 59 | 134,166 | 14,833 | 262 | 2,274.0 | 251.4 | 4.4 | 13.6% | 5.1% |
| All types | 16,968 | 873,652 | 1,930,779 | 419,919 | 51.5 | 113.8 | 24.7 | 13.0% | 5.7% |

Panel B: Addresses that redeposit borrowed tokens only

| Redeposits only | Number of addresses | Total COMP activities | | | Average COMP activities | | | % yield farming | |
| --- | --- | --- | --- | --- | --- | --- | --- | --- | --- |
|  |  | Claimed | Internal | External | Claimed | Internal | External | Internal | External |
| Large address | 66 | 385,665 | 574,021 | 229,457 | 5,843.4 | 8,697.3 | 3,476.6 | 36.4% | 22.7% |
| Small address | 2,435 | 155,673 | 273,961 | 50,917 | 63.9 | 112.5 | 20.9 | 18.6% | 7.6% |
| Yield aggregator | 2 | 613 | 0 | 0 | 306.5 | 0.0 | 0.0 | 0.0% | 0.0% |
| On-ramp | 1 | 547 | 0 | 0 | 546.9 | 0.0 | 0.0 | 0.0% | 0.0% |
| Decentralized exchange | 9 | 4,245 | 439 | 0 | 471.7 | 48.7 | 0.0 | 22.2% | 0.0% |
| Unidentified contract | 38 | 119,962 | 11,033 | 233 | 3,156.9 | 290.3 | 6.1 | 10.5% | 2.6% |
| All types | 2,551 | 666,705 | 859,453 | 280,607 | 261.4 | 336.9 | 110.0 | 19.0% | 7.8% |

**4.3 Liquidations**

As discussed in Section 2.3, Compound manages its credit risk by allowing third party users to partially liquidate loans with negative account liquidity. Figure 6 plots daily loan liquidations across all cTokens against ETH price. During periods of rapid price declines, liquidations increase. There were three major episodes of liquidations: November 16, 2020; February 22 – 23, 2021; and May 19 – 23, 2021. Daily liquidations reached USD 130 million on May 19, 2021, as the market began to crash. However, liquidations on that day represented only 2.1% of outstanding loans as evident in Figure 7 Panel B as Compound requires users to maintain ample collateralization and liquidation is only partial.

Table 8 summarizes the loans by token and address type. We require that loans be at least 100 USD so that liquidation can be economically viable with gas cost. Of the 44,423 loans, 6.1% are liquidated, representing less 0.75% of loans drawn. Liquidated stablecoin loans account for 89% of the dollar value, and liquidations of large and small addresses account for 98.3% of all liquidations. Despite their higher frequency, the dollars of liquidated loans drawn by yield aggregators and unidentified contracts account for less than 0.1% of loan value. If they primarily follow the long-short, liquidation-free strategy described in Section 2.4, then this would explain the low liquidation.

Next, we analyze the relationship between the tokens used as collateral and tokens drawn as loans. Table 9 reports the dollar values of the 5,036 liquidations that occurred between May 2019 and June 2021 by loan-collateral pairs. 86.4% of liquidated loans are in stablecoins and mostly collateralized by cETH and cWBTC, as the two tokens account for 73.9% of collaterals foreclosed by liquidators. While the implication of the long-short arbitrage strategy is such that loans exclusively collateralized by the same token cannot be liquidated, borrowers may have posted deposits in multiple tokens as collateral, and Compound allows liquidators to choose the cToken collateral they wish to receive. This is why diagonal entries such as DAI-cDAI – which means liquidated DAI loans collateralized by cDAI – are not empty.

The results highlight the importance of asset-liability price mismatch risk between collateral and borrowed tokens. Conceptually, the multi-collateral design should reduce the volatility of collateral value and borrowers could in principle delta-hedge their positions, but because most tokens are positively correlated (recall that the correlation between ETH and other tokens are between 0.73 and 0.92) especially during market downturns, liquidations are more likely to occur during such periods.

Liquidations can lead to fire sales spiral, which has occurred in many financial settings as summarized by Schleifer and Vishny (2011) and documented in DeFi by Saengchote et al. (2022). Aramonte et al. (2021) show that following forced liquidations of DeFi derivatives positions and loans, sharp price falls and spikes in volatility follow. Given that leverage is procyclical, liquidation risk can amplify instability of the DeFi ecosystem. In the next section, we discuss how the connectivity between DeFi protocols can be a source of systemic risk and how participants that are part of the events that unfolded from May 2022 interacted with Compound in February, prior to the crash.

**Figure 6: Compound liquidation.**
In this figure, daily loan liquidations in Compound are plotted against ETH price. In Panel A, the dollar value in USD millions of liquidations is plotted. In Panel B, liquidations are computed as percentage of outstanding loans.

Panel A: Liquidation in US dollars

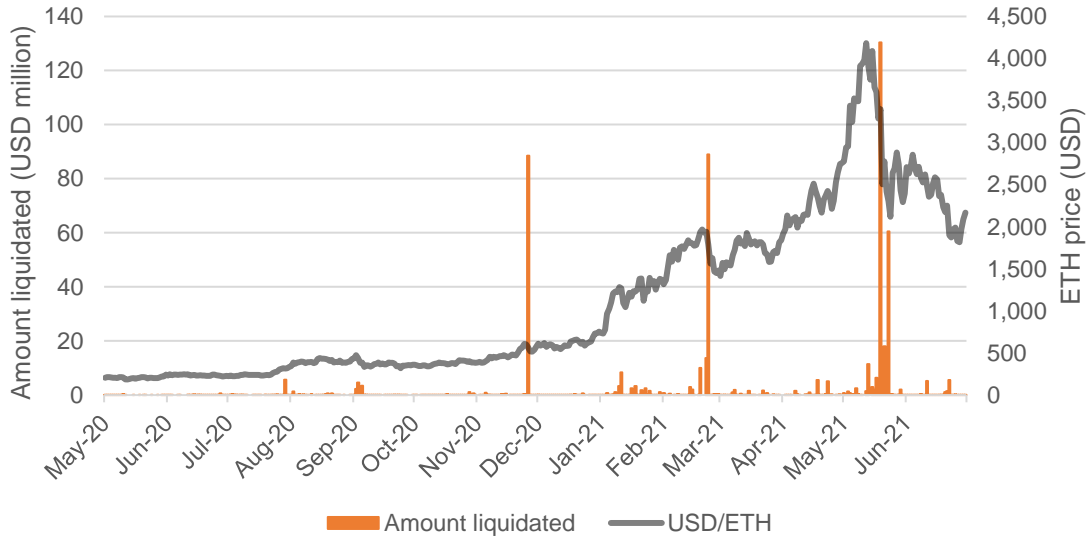

Panel B: Liquidation as percentage of outstanding loans

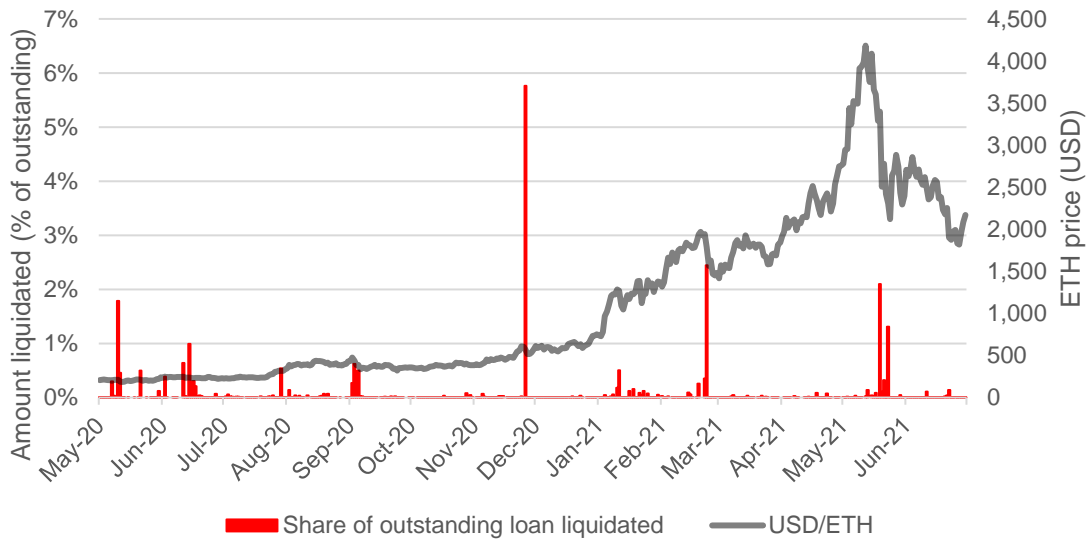

**Table 8: Liquidation statistics.**
Panel A reports the number of outstanding and closed loans as of June 2021 by tokens. Loans are required to be at least 100 USD so that liquidation can be economically viable with gas cost. The number of loans and their dollar value, the dollar value of loans repaid by liquidators (liquidation), and the proportions of loans liquidated in frequencies and dollars are reported. Panel B reports similar statistics by address type.

Panel A: Liquidation by token

| Token | Num loans | Loan value (USD m) | Liquidation (USD m) | % of loans liquidated | % of USD liquidated |
|---|---|---|---|---|---|
| ETH | 5,681 | 2,796.1 | 6.9 | 6.0% | 0.2% |
| WBTC | 1,738 | 1,900.6 | 26.3 | 3.2% | 1.4% |
| BAT | 1,973 | 962.9 | 1.7 | 3.9% | 0.2% |
| COMP | 278 | 215.8 | 8.1 | 5.8% | 3.7% |
| LINK | 138 | 86.5 | 0.0 | 0.7% | 0.0% |
| REP | 345 | 275.4 | 0.0 | 10.1% | 0.0% |
| UNI | 524 | 581.1 | 13.2 | 4.0% | 2.3% |
| ZRX | 585 | 86.7 | 0.4 | 5.6% | 0.5% |
| DAI | 14,055 | 31,693.0 | 246.9 | 7.3% | 0.8% |
| USDC | 12,233 | 24,923.5 | 115.0 | 6.5% | 0.5% |
| USDT | 6,703 | 5,218.0 | 95.8 | 4.8% | 1.8% |
| TUSD | 170 | 96.7 | 0.1 | 5.3% | 0.1% |
| All tokens | 44,423 | 68,836.2 | 514.4 | 6.1% | 0.7% |

Panel B: Liquidation by address type

| Address type | Num loans | Loan value (USD m) | Liquidation (USD m) | % of loans liquidated | % of USD liquidated |
|---|---|---|---|---|---|
| Large address | 1,383 | 30,345.1 | 210.1 | 5.7% | 0.7% |
| Small address | 42,005 | 22,696.8 | 295.6 | 6.3% | 1.3% |
| Yield aggregator | 36 | 3,309.9 | 0.4 | 13.9% | 0.0% |
| On-ramp | 2 | 256.6 | 0.0 | 0.0% | 0.0% |
| Decentralized exchange | 7 | 3.8 | 0.0 | 0.0% | 0.0% |
| Asset management | 176 | 888.8 | 0.3 | 0.6% | 0.0% |
| Unidentified contract | 814 | 11,335.2 | 8.1 | 1.5% | 0.1% |
| All types | 44,423 | 68,836.2 | 514.4 | 6.1% | 0.7% |

**Table 9: Liquidated loans by collateral.**
This table reports the total dollar value in USD millions of loans repaid by liquidators between May 2019 and June 2021. The rows are denominations of liquidated cToken loans, while the columns are the corresponding cToken collaterals seized. For example, the DAI row and cETH column represents USD 124.9 million DAI loans repaid where liquidators choose to claim cETH as collateral.

|  | cETH | cWBTC | cBAT | cCOMP | cLINK | cREP | cUNI | cZRX | cDAI | cUSDC | cUSDT | cTUSD | Loan | Share |
|---|---|---|---|---|---|---|---|---|---|---|---|---|---|---|
| ETH |  | 1.1 | 0.1 | 0.1 | 0.0 | 0.0 | 0.1 | 0.2 | 2.8 | 2.3 | 0.0 |  | 6.5 | 1.2% |
| WBTC | 36.7 | 7.0 | 0.0 | 0.0 |  |  | 0.0 | 0.0 | 0.1 | 0.2 | 0.0 |  | 44.1 | 7.9% |
| BAT | 0.4 | 0.6 |  | 0.0 |  |  | 0.0 | 0.0 | 0.0 | 2.8 |  |  | 3.9 | 0.7% |
| COMP | 0.0 | 0.7 | 0.0 | 7.1 |  |  |  |  | 0.3 | 0.0 |  |  | 8.1 | 1.4% |
| LINK | 0.0 |  |  |  |  |  |  |  | 0.0 | 0.0 |  |  | 0.1 | 0.0% |
| REP | 0.0 |  |  |  |  |  |  | 0.0 | 0.0 | 0.0 |  |  | 0.0 | 0.0% |
| UNI | 0.6 | 11.1 |  | 0.2 |  |  |  | 0.0 | 0.0 | 1.5 |  |  | 13.3 | 2.4% |
| ZRX | 0.1 | 0.0 | 0.1 | 0.0 |  |  |  |  | 0.1 | 0.0 |  |  | 0.3 | 0.1% |
| DAI | 124.9 | 15.5 | 2.9 | 2.3 |  | 0.0 | 4.0 | 0.7 | 65.9 | 17.9 | 0.0 |  | 234.3 | 41.8% |
| USDC | 99.2 | 13.0 | 1.0 | 0.6 |  | 0.0 | 1.2 | 1.2 | 11.7 |  | 0.0 |  | 128.0 | 22.8% |
| USDT | 92.1 | 11.3 | 0.7 | 0.9 |  |  | 6.6 | 5.1 | 0.5 | 4.9 |  |  | 122.0 | 21.8% |
| TUSD | 0.0 | 0.0 | 0.0 | 0.0 |  |  | 0.0 |  | 0.0 |  |  |  | 0.1 | 0.0% |
| Collateral | 354.2 | 60.3 | 4.7 | 11.1 | 0.0 | 0.0 | 12.1 | 7.2 | 81.4 | 29.7 | 0.0 | 0.0 | 560.7 | 100.0% |
| Share | 63.2% | 10.8% | 0.8% | 2.0% | 0.0% | 0.0% | 2.2% | 1.3% | 14.5% | 5.3% | 0.0% | 0.0% | 100.0% |  |

## 5. Concentration, Connectivity, and Systemic Risk

As discussed in the Introduction, one of the hallmarks of DeFi is that it is built to allow interoperability. Compound did not create DAI, but it can incorporate the stablecoin into its cDAI contract as a main component. This interoperability can be visualized by the flows between addresses. We focus on stablecoins since they are closest to money and can pose a threat to the monetary system (Arner et al., 2020), and to limit the complexity of the network diagram. We aggregate the transfers of DAI, USDC, and USDT between smart contracts during the sample period and color code the contracts (nodes) and token flows (lines).

**Figure 7: Network diagram of DeFi stablecoins.**
This figure plots the stablecoins flows (DAI, USDC, and USDT) between smart contracts identified as part of Compound, Aave (a lending protocol), yield aggregator protocols, on-ramps, decentralized exchange (DEX) protocols, asset management protocols, and other smart contracts that exist between May 2019 and June 2021.

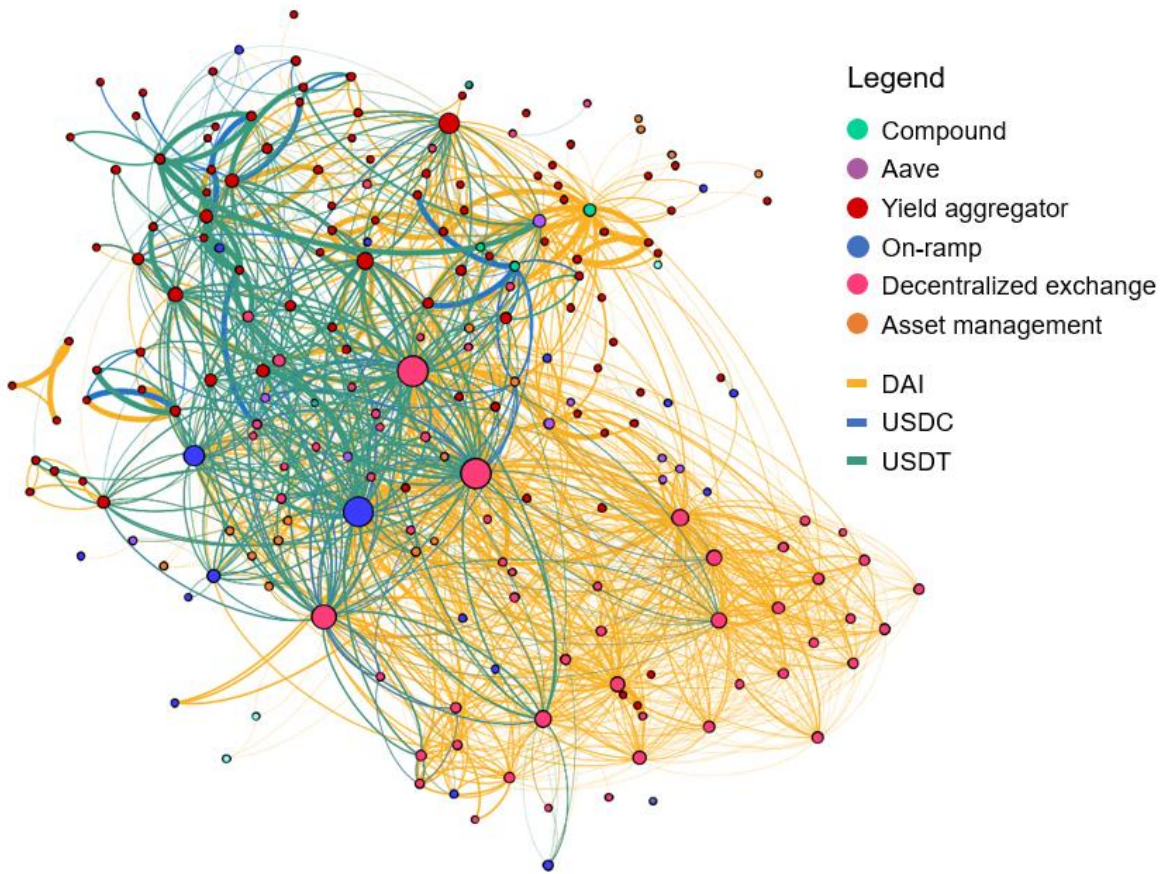

As explained in Figure 1, when a user makes a deposit into a cToken contract, she receives a cToken in return as depository receipt. The cTokens function as bearer instruments that allow whoever that possesses the tokens to redeem for corresponding tokens from the contracts. This makes the cTokens valuable and can be considered a new asset, often referred to as interest-bearing tokens, as discussed in the Appendix for yield aggregator. Because of this, they are also accepted by other DeFi protocols. For example, Curve.fi's crvCOMP contract allows cDAI and cUSDC to be exchanged in the swap pool. By depositing tokens into a protocol, users would often receive another token issued by the protocol as depository receipt. But, as seen in Figure 1, tokens can represent liabilities of the issuing contracts, so the chains of interactions illustrated by Figure 7 is similar to the debt-on-debt network of financial liabilities that stem from the original asset, such as ETH that secures WETH, which, in turn, secures cWETH, and so on.

Dang et al. (2020) argues that short-term debt designed to be information insensitive (such as bank deposits or T-bills) can serve as a medium of exchange. In the context of DeFi, certain classes of stablecoins are examples of such short-term debt. When there is doubt about the value of the collateral backing the short-term debt, users may begin acquiring private information, shifting the state from information-insensitive debt to information-sensitive debt, which can in turn result in financial crises when then credit chain collapses. Saengchote (2021b) demonstrates that composability in DeFi could lead to tacit leverage via webs of depository receipt creation that is not easy to recognize or monitor, on-chain lenders may provide explicit leverage that further amplify the financial connectivity in the ecosystem. When the price of the initial collateral declines, the whole credit chain will be affected.

In May 2022, the failure of TerraUSD stablecoin had a ripple effect throughout the crypto asset market, with Bitcoin falling to its lowest price since 2020.[28] Several crypto businesses such as Three Arrows Capital (a crypto hedge fund), Celsius Network (a crypto bank), and Voyager Digital (a crypto brokerage / bank) filed for bankruptcy in the ensuing months.[29] The effect of the May collapse continued to reverberate, and in November, FTX (one of the world's biggest crypto exchange) filed for bankruptcy and Alameda Research (a crypto hedge fund) that is related FTX was reported to be insolvent.[30] All of the examples earlier are not purely DeFi providers, as not all of their activities are completely on-chain, but because ultimately they interact with DeFi protocols, on-chain data can provide some insights into what was happening before the May crash.

---

[28] Source: https://www.nytimes.com/2022/05/12/technology/cryptocurrencies-crash-bitcoin.html, accessed November 28, 2022.
[29] Source: https://www.businessinsider.com/why-crypto-celsius-three-arrows-voyager-filed-bankruptcy-2022-7, accessed November 28, 2022.

Before we proceed further, we remind the reader that the analysis that follows is an attempt in an earlier version of our paper to investigate the concentration risk of DeFi by looking at who are the main depositors and borrowers in Compound prior to May. Many of the addresses that were anonymous to us on February 11, 2022, have now been identified by the DeFi community and various familiar names emerge. We report the top 10 holders of cETH, cDAI, cUSDC, and cUSDT as of June 30, 2021, and February 11, 2022, in Table A3, and the top 20 borrowers over the sample period in Table A4.

On February 15, 2022, the-then pseudonymous, independent investigative journalist named Dirty Bubble Media released a publication titled "Celsius Network's unsustainable DeFi strategy could be costing them millions".[31] Celsius Network allows users to deposit their crypto assets and earn yield. While it does not disclose the sources of its revenue, DeFi protocols are potential sources, and on-chain data can provide some insights. The author analyzed the yield generated from Compound and Aave and raised questions regarding its ability to generate revenue and long-term viability:

> *"These deposits generate some interest. However, you'll note that the APY for these deposits is quite low; lower, in fact, than many "tradfi" savings accounts. This means there is a significant gap between what Celsius is paying and what they are receiving as interest on these deposits. Based on a conservative estimate of the average APY Celsius offers customers on these crypto assets, they face an annual deficit of ~$86 million in interest payments to depositors.*

The author then released a follow-on publication, highlighting that investments in ETH and related crypto assets account for more than 65% of its assets under management, most of which are held by a single address.[32] The potential vulnerability of Celsius Network's business model, its concentrated position in certain crypto assets, and potential contagion that can occur from fire sales spiral are ingredients for systemic crash that can be hard to contain given most providers are unregulated and no backstops to prevent runs like those studied by Diamond and Dybvig (1983) exist.

---

[30] Source: https://www.nytimes.com/2022/11/11/business/ftx-bankruptcy.html, accessed November 28, 2022.
[31] Source: https://dirtybubblemedia.substack.com/p/celsius-networks-unsustainable-defi, accessed November 28, 2022.
[32] Source: https://dirtybubblemedia.substack.com/p/following-the-money, accessed November 28, 2022.

In Panel A of Table A3, the top holder of cETH both in June 2021 and February 2022 were indeed Celsius Network, and it is the same address as reported in the publication. In fact, for all four tokens, the concentration of top 10 holders increased over the 7.5 months, and for cDAI, the top 10 addresses account for almost 90% of the deposits. In addition, yield aggregators became more prevalent in February 2022, with Yearn becoming the dominant protocol and is the majority holders of cDAI (21.8%) and cUSDC (11.7%). Moreover, among the top 10 holders is a lending protocol (Notional Finance) that accepts cTokens (cETH, cWBTC, cDAI, and cUSDC) as collateral. These protocols increase the connectedness among participants in DeFi.

Further investigation of cumulative borrowing activities up to June 2021 reported in Table A4 shows that the distressed names such as Alameda Research and Three Arrows Capital are among the top 10 borrowers. The address marked as Three Arrows Capital claimed more than 27,000 COMP reward during the sample period, suggesting that they participated in some degree of leveraged yield farming within Compound, while the Alameda Research address has cumulative net withdrawals of USD 6.7 billion, which could result from buying cTokens in secondary markets to redeem for deposits, which, depending on market prices, could provide arbitrage profits. Because of blockchain's pseudonymity, it is possible that other unidentified addresses belong to these names, or some other crypto businesses. The figures are as of June 2021; by May 2022, their activities could have grown much more from another rise of the crypto asset market in late 2021.

In traditional finance, such concentration risk would be curtailed by risk management practices such as borrowing limits and identification of consolidated debt positions across related parties. However, with the pseudonymity of DeFi, such practices are very difficult. While not all their transactions are completely on-chain (hence such crypto businesses are often referred to as "CeDeFi", a combination of centralized and decentralized finance), ex-post investigations such as from bankruptcy filings reveal the credit relationships among these entities, making their interconnection even tighter than on-chain data suggests.[33] Acemoglu et al. (2015) show that dense interconnections propagate shocks rather than enhance financial stability, and on-chain data shows that this is indeed the case for DeFi.

---

[33] For example, the Chapter 11 petition filed by Celsius Network on July 14, 2022, shows that Alameda Research is its one of its top 50 creditors. Source: https://pacer-documents.s3.amazonaws.com/115/312902/126122257414.pdf, accessed November 28, 2022.

Prior to the collapse of FTX, The FSOC issued a Report on Digital Asset Financial Stability Risks and Regulation in early October 2022, citing "Financial Exposures via Interconnections within the Crypto Asset Ecosystem" as a potential source of risk.

*Interconnections inside the crypto-asset ecosystem spread losses if a shock causes the default of an interconnected entity and its counterparties then incur knock-on losses. Losses can also spread from common holdings if an entity holds a crypto-asset that records a sharp price decline. The crypto-asset ecosystem currently features a number of significantly interconnected entities, including crypto-asset platforms, investors, and other counterparties. The failure of a significantly interconnected entity can cause substantial distress within the crypto-asset ecosystem.*

Our investigations are conducted with hindsight, and we do not claim that the collapses of the crypto businesses discussed earlier could have been predicted or prevented, since many addresses identified in this paper were previously unrecognized to many until after the collapse. However, blockchain transparency can inform us about the nature of interactions that occur within DeFi. When much of the activities are leveraged, yield-chasing strategies that are supported by risky collateral and concentrated among few large participants, the credit chain can be very fragile.

## 6. Conclusion

In this paper, we outline how Compound – a DeFi lending/money market protocol – works, who its users are, and how they interact with the protocol. Between May 2019 and June 2021, $61.4 billion of loans in multiple tokens are made for an average duration of 31 days, with usage behavior consistent with leveraged investment strategy rather than financing. We document some novel facts of DeFi, such as users creating addresses and interacting in small amounts (the micro addresses) in anticipation of COMP airdrop (which did not eventually occur), and COMP rewards remaining unclaimed because of gas cost and real option value.

Depositors and borrowers in Compound are concentrated, with the top 100 depositor addresses accounting for 75% of all deposits and the top 100 borrower addresses 78% of all loans, while the top 10 holders of selected cTokens accounting for 50 to 90% of outstanding tokens. We find that deposit demands are largely idiosyncratic and do not respond to returns to deposits and market conditions, while loan demands are driven by leveraged yield farming and are responsive to COMP rewards. Further address-level and loan-level investigations reveal results that are consistent with this interpretation, with 11.4% of drawn loans redeposited to the protocol again within 1 day, and addresses that are more likely to employ this strategy claiming more COMP rewards from Compound. We also find that the users interact with different tokens in different ways, even for stablecoins that are supposed to trade at price close to USD, showing that each token have different roles in the DeFi ecosystem.

As highlighted by The Financial Stability Oversight Council (FSOC), interconnections between different protocols on-chain and different crypto businesses off-chain makes the crypto asset ecosystem prone to systemic risk. While DeFi protocols such as Compound have strict credit risk management mechanisms via overcollateralization, other protocols can be built with less strict credit policies. Protocols that rely on stablecoin collaterals tend to be more generous with required level of collateralization but the failure of TerraUSD stablecoin shows that stablecoins can also sharply lose their values. And as discussed earlier, certain blockchain oracles can be manipulated to overstate collateral values. And when that happens, the seemingly safe loans can also be liquidated. Aramonte et al. (2021) show that automatic liquidation mechanisms can amplify shocks and instability of the financial system.

It is now clear that lending protocols can contribute to systemic risk,[34] and events from May to November 2022 have shown that significant vulnerabilities can also build up in the off-chain crypto businesses that are intertwined with on-chain DeFi protocols. At present, the regulatory guardrails against excessive leverage for participants in the crypto asset ecosystem are not uniform, and DeFi regulations can be difficult to implement given how permissionless blockchains are designed to operate. In addition, with the currently ambiguous status of crypto assets and the global nature of crypto businesses, uncertainty regarding who ultimately has jurisdiction remains. This is the challenge that can determine the success or failure of the crypto asset ecosystem which ultimately has nothing to do with the underlying technology, as it is always the people who use the technology that matter.

---

[34] In previous versions that were circulated since September 2021, we stated that "yield farming warrants further investigation into whether DeFi composability lends itself to systemic risk". Ex-post investigations by various sources such as the FSOC have now confirmed that many crypto businesses that collapsed were taking part in some forms of leveraged yield farming.

**APPENDIX: Smart contract and DeFi protocol classification**

In this appendix, we briefly explain the nature of DeFi protocols and how their incentives are distributed so readers can understand the distinctions and the reason behind the classification scheme.

1. **Yield aggregator**

Yield aggregator protocols are similar to mutual funds and hedge funds. In programmable blockchains, users are required to pay gas to compute and record new information on to the blockchain regardless of the monetary value behind that information because decentralized computing power and block space are limited resources in such blockchains. Claiming reward tokens (such as COMP) is a blockchain transaction which costs gas, so users with small transactions will not find it economical to claim rewards often, missing out on the compounding effect. With larger pool of tokens, yield aggregators can claim rewards more frequently, and thus accumulate more yield overtime. In addition, yield aggregators can utilize complex strategies such as recursive leverage or staking wrapped tokens across multiple protocols to earn rewards distributed by those protocols. In return, the protocol will charge management fees on assets under management may also withhold a proportion of the yield (like hedge fund carry). Some strategies are illiquid, so mechanisms such as load fees are often included in protocol design to encourage users to lock their funds for longer. Examples of such protocols are Akropolis, Alpha Homora, Harvest, Idle and Yearn.finance.

Users deposit their tokens into the protocol's vault, creating a wrapped tokens as depository receipts. These tokens are often referred to as interest-bearing tokens and the small alphabet(s) in front of the original token's names depict this relationship. For example, cDAI and aDAI can be considered Compound's and Aave's interest-bearing tokens, while Yearn will create yvDAI when DAI is deposited into the DAI yVault. The deposited tokens would then be deployed according to the strategies set forth by the protocol. When the depository receipts are redeemed, users would get back a pro-rata share of pool. Because the depository receipts are also tokens, they are tradeable and can be further deposited into protocols that accept them.

Yield aggregators may also form partnerships with other protocols. For example, Yearn created its governance token YFI in July 2020, but it would only be distributed to users who staked their yTokens (Yearn's interest-bearing tokens received from depositing stablecoins) in Curve's liquidity pool. This type of interoperability is possible if smart contracts grant permissions to interact with one another.

Yield aggregator protocols identified via manual inspection of top addresses in this paper are: 88mph, DeFi Saver, Furucombo, Harvest, Idle, Inverse, Mushroom, PoolTogether, Rain Capital, Robo, Shell, Volatility and Yearn.

2. **On-ramp service providers**

On-ramp service providers are addresses that identify themselves as belonging to centralized exchanges such as Binance and Coinbase, as well as semi-centralized service such as InstaDapp. They aggregate orders and transact on behalf of clients, providing access points to the DeFi ecosystem.

On-ramp service providers identified via manual inspection of top addresses in this paper are: Binance, Dharma Finance, Eth2Dai, and InstaDapp.

## 3. Decentralized exchange (DEX)

Decentralized exchange protocols are sometimes referred to as automated market maker (AMM) protocols as they facilitate token exchanges without the need for a centralized, off-chain institution that typically uses order book matching system. The inherent reason why order book matching in DeFi is not popular is because order flows generate data trails that are extremely costly to record on the blockchain. As such, unmatched orders submitted on-chain would still cost gas, necessitating an alternative design.

Just as the name AMM suggests, participants are, in fact, market makers who must then face inventory risk. Users who provide liquidity in a pool by depositing tokens become willing counterparties for users who wish to exchange their tokens. In the order book matching system, users send in the desired orders, which are then matched to counterparties with the same terms of trade, providing price certainty at the expense of execution uncertainty. In AMM, users send one type of token she wishes to exchange, and the pool will send the other type of token in return. A bonding curve (which is essentially a pricing function) will determine how many tokens of the other type she will receive. In other words, the user will have execution certainty (provided that she pays enough gas, and the pool has sufficient liquidity to exchange) but faces price uncertainty since price is a mathematical output of the bonding function. The act of exchanging is often referred to as a swap.

The bonding curve is a function of quantities of tokens available in the pool, so large transactions *will* result in price slippage. A self-stabilizing bonding curve should generate relative token prices that make the token type in low supply prohibitively expensive to acquire (and vice versa). When the relative price in a pool is different from other trading venues, arbitrageurs can act to restore price parity. One popular example of a bonding curve is the constant product function ($xy = k$) where $x$ and $y$ are quantities of two tokens in a pool and $k$ is a constant. For given quantities $x$ and $y$, the exchange rate is the ratio of the two tokens. This function was first suggested by Vitalik Buterin (co-founder of the Ethereum blockchain) in a blog post in 2016,[35] and further developed by Hayden Adams into Uniswap protocol in 2018.

Lehar and Parlour (2021) show how a small pool can have wild swings in prices that are out-of-sync with other trading venues, but the price difference diminishes as pool size grows, but Park (2021) argues that it gives rise to sandwich attacks similar to front-running that that increase the cost of trading and threaten the long-term viability of the DeFi eco-system.

---

[35] Source: https://www.reddit.com/r/ethereum/comments/55m04x/lets_run_onchain_decentralized_exchanges_the_way/, accessed on November 11, 2022.

Despite these shortcomings, with its simplicity, the constant product function is the most popular and adopted by many protocols such as Uniswap and SushiSwap. However, it permits only a pair of tokens. More generalized versions of bonding curves can allow for more tokens, such as the more complex constant functions of Balancer or Curve, the latter of which is designed for and is the most favored protocol for stablecoin swaps. For some specifications of bonding curves, as market prices change, the ratios of tokens in liquidity pools will change to keep up with market prices. Consequently, some DEX protocols can also be viewed as asset management protocol that provides automatic portfolio rebalancing for users without paying gas.

As protocol performance directly depends on liquidity, protocol developers often provide generous rewards for users willing to provide liquidity, especially when the market is thin, leading to the term "liquidity mining" being used to describe this activity. In fact, some of the most generous rewards are often found in the nascent days of a DEX protocol as it tries to attract liquidity.[36] Rewards could be provided in the protocol's native tokens, or other protocol's native tokens if a partnership between protocols can be formed. For example, the Aave liquidity pool on Curve which accepts aDAI, aUSDC, and aUSDT distributes CRV (Curve's governance token) and stkAAVE (staked version of AAVE) as rewards.

Like yield aggregators, when tokens are deposited, users receive depository receipts, which in this context are referred to as LP tokens, representing pro-rata share of the pool. Most DEX pools accrue transaction fees, so users will also get their share of fees upon redemption. For liquidity pools containing tokens with low price correlation, the value of the redeemed pro-rata tokens can be different from the value of the deposited tokens. This small discrepancy arises from the curvature of the bonding curve and the severity depends on how the prices of the tokens diverge, leading to this counterfactual loss being called "divergent loss" or "impermanent loss". For pools where tokens prices are highly correlated such as stablecoins or wrapped tokens, this problem is less pertinent; consequently, such pools also tend to receive little or no liquidity mining reward.

The reliance on the depth of liquidity is not limited to DEX but a general feature of peer-to-pool transactions in DeFi. As discussed earlier, lending protocols also require liquidity (reflected in utilization ratio), otherwise interest rates will adjust. This highlights the nature of DeFi that smart contracts can be viewed as another form of intermediary with explicit rules governing interactions between users and the intermediary. The rules are hardcoded into blockchain and enforced autonomously without discretion or prejudice. Because a protocol's efficiency is directly influenced by the protocol's liquidity, developers often employ strategies such as offering liquidity mining rewards to bootstrap liquidity from users.

Decentralized exchange protocols identified via manual inspection of top addresses in this paper are: 1inch, BlackHoleSwap, Curve, ParaSwap, SushiSwap and Uniswap

4. **Asset management**

---

[36] For example, Uniswap only provided liquidity mining reward for two months in 2020. Source: https://www.theblockcrypto.com/linked/84762/dex-uniswap-liquidity-mining-over, accessed on July 26, 2021.

While the services provided by yield aggregators can also be considered asset management, in this paper we define asset management protocols as protocols that behave like indexed funds, users to maintain a balanced exposure to a basket of tokens or a specific strategy. This is often done by creating a pool of tokens where the proportion of each token mirrors an index, where depositors, in turn, are given depository tokens like fund units. Because the units are tradeable assets, this makes them more similar to exchange-traded funds (ETFs) than mutual funds, although the creation and redemption process is open to all, not limited to authorized market participants. Examples of such protocols are Set Protocol and Balancer.

There are few protocols under this category because some liquidity pools in decentralized exchanges can also be considered asset management protocols, maintaining a balanced exposure to the tokens they hold, but the permissible baskets are much more limited (e.g. two tokens only, or stablecoins only) and the exposure rule is defined by the bonding curve rather than an index. Under this definition, stablecoin protocols that create stablecoins from other stablecoins (e.g. mStable) can also be considered an asset management protocol, as it is indexed to the value of US dollar.

Asset management protocols identified via manual inspection of top addresses in this paper are: BasketDAO, DeFiner, Index Coop, Origin Dollar, PieDAO, Set Protocol and mStable.

### 5. Unidentified contracts

Ethereum addresses that have codes written inside are classified as smart contracts rather than wallets. This information is visible on blockchain explorer websites such as Etherscan.io. However, not all smart contracts disclose their source codes and their affiliations, and all we can see is their binary data. One example is '0x0000006daea1723962647b7e189d311d757Fb793' which, as of July 26, 2021, holds records of over 546,400 transactions and 124 types of tokens worth over $104 million. However, nothing else about the address is known. Nevertheless, not all contracts are as active and valuable as this example.

Complicated instructions such as recursive borrowing (also employed by yield aggregator protocols) can be automated via smart contracts. These addresses may belong to individuals or institutions but suggest a higher level of sophistication than simple addresses that need to submit each step of the instructions manually, so we separate them into a distinct classification. It is also possible that they are part of a yield aggregator protocol, but are not labeled explicitly in Etherscan.io, making them harder to identify.

**APPENDIX: Determinants of Net Deposits and Loan Demand by Address Type**

**Table A1: Determinants of net deposits by address type.**

This table reports the result from the regressions of daily net deposits (deposits minus withdrawals) in USD million between May 2019 and June 2021 for the 5 main tokens: ETH, WBTC, DAI, USDC, and USDT. Similar to Table 3, the explanatory variables include lagged supply rate, its interaction with an indicator for periods post COMP distribution, lagged supply COMP reward rate, log dollar value of deposits in USD million, 1-day ETH return, 7-day ETH return, and 30-day ETH volatility (measured in percentage point). The flows to and from are aggregated by address type. Panel A reports the results for large address, Panel B for small address, Panel C for yield aggregator and asset management protocols, Panel D for unidentified contract, and Panel E for addresses that also borrow. Standard errors are computed using the Newey-West procedure with one-day lag and reported in parenthesis. Stars correspond to statistical significance level, with *, ** and *** representing 10%, 5% and 1% respectively.

Panel A: Large address

|  | (1) cETH | (2) cWBTC | (3) cDAI | (4) cUSDC | (5) cUSDT |
|---|---|---|---|---|---|
| Lagged supply rate | -9.46 | 0.151 | -5.81 | -1.97 | 0.588 |
|  | (16.79) | (0.24) | (3.56) | (2.11) | (0.51) |
| Post COMP distribution | -1.71 | 2.90 | -13.74 | -12.21 | 0.719 |
|  | (4.82) | (2.26) | (16.33) | (12.57) | (1.39) |
| Post * lagged supply rate | 27.22 | 3.14 | 6.45 | 2.17 | -0.236 |
|  | (53.96) | (4.18) | (4.17) | (3.19) | (0.55) |
| Lagged supply reward | 5.17 | -2.65 | 0.192 | 7.49 | 0.016 |
|  | (12.38) | (3.09) | (1.82) | (5.48) | (0.23) |
| Log(pool size) | -0.002 | -0.001 | -0.004 | -0.004 | -0.008* |
|  | (0.000) | (0.000) | (0.000) | (0.010) | (0.000) |
| ETH return (1d) | 6.57 | -21.54 | -63.84 | -68.13 | -30.28* |
|  | (54.21) | (28.62) | (83.09) | (120.52) | (16.18) |
| ETH return (7d) | -3.03 | 14.79 | 29.92 | -4.04 | 7.45 |
|  | (22.79) | (13.92) | (29.22) | (59.90) | (6.44) |
| ETH SD (30d) | -37.08 | 1.955 | 110.59 | -237.08 | 75.38* |
|  | (230.14) | (117.53) | (220.28) | (415.71) | (44.93) |
| Constant | 2.76 | -1.29 | 14.14 | 12.71 | -4.49** |
|  | (9.24) | (4.23) | (12.54) | (15.91) | (1.94) |
| Observations | 426 | 351 | 426 | 426 | 425 |
| Adj R-squared | -0.014 | -0.017 | 0.003 | -0.006 | 0.037 |

Panel B: Small address

|  | (1) cETH | (2) cWBTC | (3) cDAI | (4) cUSDC | (5) cUSDT |
|---|---|---|---|---|---|
| Lagged supply rate | -12.83 | 1.64*** | 0.213 | -0.202 | 1.29** |
|  | (8.83) | (0.100) | (1.31) | (0.390) | (0.550) |
| Post COMP distribution | -4.47 | 2.64*** | -2.16 | -3.29* | 3.24** |
|  | (2.72) | (0.880) | (4.41) | (1.73) | (1.48) |
| Post * lagged supply rate | 22.571 | 0.362 | 0.164 | 1.27*** | -0.842 |
|  | (23.23) | (2.02) | (1.37) | (0.45) | (0.56) |
| Lagged supply reward | 1.61 | -0.957 | 0.108 | 0.047 | -0.149 |
|  | (4.63) | (1.24) | (0.420) | (0.480) | (0.140) |
| Log(pool size) | 0.000 | 0.000 | -0.001 | 0.000 | 0.001 |
|  | (0.000) | (0.000) | (0.000) | (0.000) | (0.000) |
| ETH return (1d) | -7.63 | -24.49** | 13.98 | -0.204 | 8.56 |
| | (18.70) | (12.18) | (13.12) | (12.40) | (5.69) |
| ETH return (7d) | 11.052 | 8.25 | 8.52 | -0.330 | 1.69 |
| | (9.71) | (5.27) | (7.03) | (6.28) | (2.90) |
| ETH SD (30d) | -62.77 | -75.59** | 16.16 | 59.59 | -6.85 |
| | (67.00) | (38.24) | (35.11) | (39.09) | (29.26) |
| Constant | 4.16 | 1.16 | 1.29 | -1.35 | -4.10** |
| | (3.06) | (1.27) | (2.80) | (1.76) | (1.68) |
| | | | | | |
| Observations | 426 | 351 | 426 | 426 | 425 |
| Adj R-squared | 0.001 | 0.023 | 0.003 | 0.080 | 0.126 |

Panel C: Protocols

| | (1) cETH | (2) cWBTC | (3) cDAI | (4) cUSDC | (5) cUSDT |
|---|---|---|---|---|---|
| Lagged supply rate | 3.26 | -0.006 | -0.034 | -0.551 | 0.007 |
| | (3.12) | (0.010) | (0.140) | (0.430) | (0.010) |
| Post COMP distribution | -0.806 | 0.200 | -0.910 | -1.07 | -0.058 |
| | (0.92) | (0.160) | (1.48) | (1.92) | (0.360) |
| Post * lagged supply rate | 17.79* | -0.005 | 0.311 | 0.615 | 0.001 |
| | (10.23) | (0.11) | (0.23) | (0.49) | (0.04) |
| Lagged supply reward | -3.21 | 0.085 | -0.063 | 0.260 | 0.025 |
| | (2.79) | (0.160) | (0.200) | (0.730) | (0.070) |
| Log(pool size) | -0.002*** | 0.000 | 0.000 | -0.000 | -0.000 |
| | (0.00) | (0.000) | (0.000) | (0.000) | (0.000) |
| ETH return (1d) | 5.27 | 2.03 | 18.34** | -10.04 | -8.14** |
| | (7.39) | (1.67) | (7.37) | (13.27) | (3.48) |
| ETH return (7d) | 12.18** | 0.841 | -2.53 | 2.94 | 0.511 |
| | (5.97) | (0.620) | (3.33) | (5.70) | (1.21) |
| ETH SD (30d) | 72.18** | -2.82 | -3.27 | 26.13 | -0.814 |
| | (31.67) | (3.71) | (21.02) | (45.01) | (8.20) |
| Constant | -2.93** | -0.012 | 0.162 | 0.001 | 0.039 |
| | (1.22) | (0.120) | (0.830) | (1.77) | (0.300) |
| | | | | | |
| Observations | 426 | 351 | 426 | 426 | 425 |
| Adj R-squared | 0.150 | -0.001 | 0.016 | -0.011 | 0.001 |

Panel D: Unidentified contract

| | (1) cETH | (2) cWBTC | (3) cDAI | (4) cUSDC | (5) cUSDT |
|---|---|---|---|---|---|
| Lagged supply rate | 2.28 | -0.041 | 0.021 | -1.62 | -0.164 |
| | (3.05) | (0.040) | (0.020) | (2.03) | (0.32) |
| Post COMP distribution | -0.436 | -0.212 | -0.002 | -17.93 | -0.910 |
| | (0.570) | (0.480) | (0.050) | (19.37) | (3.08) |
| Post * lagged supply rate | -5.25 | -0.520 | -0.027 | 5.44** | 0.599 |
| | (5.06) | (1.13) | (0.020) | (2.50) | (0.52) |
| Lagged supply reward | 0.572 | 0.368 | 0.009 | 1.31 | -0.812 |
| | (0.960) | (0.650) | (0.020) | (2.04) | (1.03) |
| Log(pool size) | 0.001*** | 0.000 | 0.000 | 0.001 | 0.002 |

|  | (0.000) | (0.000) | (0.000) | (0.001) | (0.000) |
| --- | --- | --- | --- | --- | --- |
| ETH return (1d) | 6.34 | -2.80 | 0.779 | 53.98 | 31.38 |
|  | (5.13) | (7.82) | (0.520) | (80.28) | (21.47) |
| ETH return (7d) | -3.11 | 2.96 | -0.519 | 58.06 | 2.39 |
|  | (2.35) | (3.52) | (0.360) | (49.03) | (11.78) |
| ETH SD (30d) | -10.81 | 41.82 | -0.313 | -262.56 | 122.56 |
|  | (17.89) | (38.26) | (1.19) | (256.86) | (94.58) |
| Constant | 0.036 | -1.44 | 0.033 | 9.86 | -4.60 |
|  | (0.800) | (1.14) | (0.060) | (10.06) | (3.48) |
|  |  |  |  |  |  |
| Observations | 426 | 351 | 426 | 426 | 425 |
| Adj R-squared | 0.095 | -0.014 | 0.030 | 0.016 | 0.022 |

Panel E: Borrowers only

|  | (1) cETH | (2) cWBTC | (3) cDAI | (4) cUSDC | (5) cUSDT |
| --- | --- | --- | --- | --- | --- |
| Lagged supply rate | -19.57 | 1.71*** | -7.15 | -2.04 | 1.08* |
|  | (19.69) | (0.240) | (5.35) | (2.26) | (0.600) |
| Post COMP distribution | -6.44 | 4.16* | -34.09 | -14.63 | 3.22** |
|  | (5.66) | (2.46) | (28.01) | (14.62) | (1.47) |
| Post * lagged supply rate | 46.35 | 2.64 | 11.39* | 3.03 | -1.098* |
|  | (59.10) | (5.24) | (6.22) | (3.48) | (0.640) |
| Lagged supply reward | 7.92 | -3.18 | 2.22 | 7.24 | 0.061 |
|  | (13.66) | (3.82) | (3.27) | (5.95) | (0.220) |
| Log(pool size) | -0.001 | 0.000 | -0.005 | -0.004 | -0.009* |
|  | (0.000) | (0.000) | (0.010) | (0.010) | (0.010) |
| ETH return (1d) | 2.10 | -30.94 | 16.85 | -35.47 | -8.48 |
|  | (54.91) | (33.48) | (134.92) | (141.49) | (9.38) |
| ETH return (7d) | 3.93 | 12.57 | 93.54 | -14.46 | 5.43 |
|  | (24.58) | (17.21) | (64.38) | (72.42) | (6.15) |
| ETH SD (30d) | -108.140 | -23.34 | -105.32 | 27.744 | 94.81* |
|  | (239.86) | (144.52) | (411.14) | (495.63) | (50.14) |
| Constant | 6.94 | -0.824 | 25.32 | 3.91 | -6.83*** |
|  | (9.73) | (5.08) | (19.89) | (18.99) | (2.28) |
|  |  |  |  |  |  |
| Observations | 426 | 351 | 426 | 426 | 425 |
| Adj R-squared | -0.013 | -0.018 | 0.013 | -0.011 | 0.034 |

**Table A2: Determinants of loans drawn by address type.**
This table reports the result from the regressions of daily net deposits (deposits minus withdrawals) in USD million between May 2019 and June 2021 for the 5 main tokens: ETH, WBTC, DAI, USDC, and USDT. Similar to Table 4, the explanatory variables include lagged borrow rate, its interaction with an indicator for periods post COMP distribution, lagged borrow COMP reward rate, log dollar value of deposits in USD million, 1-day ETH return, 7-day ETH return, and 30-day ETH volatility (measured in percentage point). The flows to and from are aggregated by address type. Panel A reports the results for large address, Panel B for small address, Panel C for yield aggregator and asset management protocols, and Panel D for unidentified contracts. Standard errors are computed using the Newey-West procedure with one-day lag and reported in parenthesis. Stars correspond to statistical significance level, with *, ** and *** representing 10%, 5% and 1% respectively.

Panel A: Large address

|  | (1) cETH | (2) cWBTC | (3) cDAI | (4) cUSDC | (5) cUSDT |
|---|---|---|---|---|---|
| Lagged borrow rate | 5.52*** | -0.079 | 0.265** | 0.332 | 0.638*** |
|  | (1.29) | (0.090) | (0.120) | (0.310) | (0.110) |
| Post COMP distribution | -6.14 | -3.37** | -2.48*** | 2.48 | 9.16*** |
|  | (5.45) | (1.70) | (0.790) | (1.80) | (1.54) |
| Post * lagged borrow rate | -0.118 | 0.327 | -0.149 | -0.423 | -0.511*** |
|  | (2.06) | (0.330) | (0.120) | (0.310) | (0.120) |
| Lagged borrow reward | 0.574*** | 0.213* | 0.257*** | 0.280*** | -0.301** |
|  | (0.120) | (0.120) | (0.050) | (0.080) | (0.150) |
| Log(pool size) | 1.15*** | 0.844*** | 1.31*** | 2.32*** | 1.41*** |
|  | (0.420) | (0.250) | (0.150) | (0.250) | (0.150) |
| ETH return (1d) | 1.90 | 0.413 | 2.81 | -1.42 | 1.58 |
|  | (5.88) | (6.68) | (2.46) | (2.31) | (2.96) |
| ETH return (7d) | 0.243 | 2.97 | -0.193 | 1.35 | -0.134 |
|  | (3.04) | (3.58) | (1.03) | (1.23) | (1.49) |
| ETH SD (30d) | -30.94 | 43.18 | -34.12*** | -65.49*** | -24.35*** |
|  | (23.11) | (31.35) | (9.46) | (9.68) | (8.42) |
| Constant | -10.00*** | -1.48 | 8.33*** | 0.392 | -1.20 |
|  | (3.11) | (1.44) | (1.07) | (1.99) | (1.14) |
| Observations | 426 | 351 | 426 | 426 | 425 |
| Adj R-squared | 0.175 | 0.095 | 0.364 | 0.596 | 0.585 |

Panel B: Small address

|  | (1) cETH | (2) cWBTC | (3) cDAI | (4) cUSDC | (5) cUSDT |
|---|---|---|---|---|---|
| Lagged borrow rate | 2.49*** | 0.198*** | 0.087 | -0.025 | 0.247*** |
|  | (0.560) | (0.070) | (0.060) | (0.080) | (0.090) |
| Post COMP distribution | 3.84** | 0.676 | -1.22*** | 0.150 | 3.30*** |
|  | (1.68) | (0.91) | (0.41) | (0.52) | (1.02) |
| Post * lagged borrow rate | -1.15* | -0.477** | -0.052 | -0.008 | -0.226*** |
|  | (0.660) | (0.220) | (0.060) | (0.090) | (0.090) |
| Lagged borrow reward | 0.015 | 0.102** | 0.143*** | 0.103*** | 0.095* |
|  | (0.040) | (0.050) | (0.030) | (0.030) | (0.050) |

|  | 0.620*** | 0.267** | 0.905*** | 0.973*** | 0.974*** |
| --- | --- | --- | --- | --- | --- |
| Log(pool size) | (0.090) | (0.130) | (0.080) | (0.070) | (0.120) |
| ETH return (1d) | 2.36* | -3.96 | 1.37 | 0.461 | -1.99 |
|  | (1.29) | (2.86) | (1.21) | (0.930) | (1.25) |
| ETH return (7d) | -0.541 | 0.203 | 0.788 | 0.525 | 1.67*** |
|  | (0.660) | (1.27) | (0.600) | (0.470) | (0.64) |
| ETH SD (30d) | -4.37 | 23.02 | -14.83*** | -17.57*** | -14.18** |
|  | (5.64) | (14.79) | (4.44) | (3.88) | (5.73) |
| Constant | 2.02 | 8.40*** | 9.88*** | 9.49*** | 6.86*** |
|  | (1.39) | (0.790) | (0.490) | (0.530) | (0.940) |
|  |  |  |  |  |  |
| Observations | 426 | 351 | 426 | 426 | 425 |
| Adj R-squared | 0.505 | 0.048 | 0.549 | 0.669 | 0.645 |

Panel C: Protocols

| VARIABLES | (1) cWBTC | (2) cUSDC |
| --- | --- | --- |
| Lagged borrow rate | 0.036 | -0.125 |
|  | (0.030) | (0.270) |
| Post COMP distribution | -3.02*** | -1.37 |
|  | (0.67) | (1.59) |
| Post * lagged borrow rate | 0.487*** | -0.140 |
|  | (0.160) | (0.280) |
| Lagged borrow reward | 0.084* | -0.277** |
|  | (0.050) | (0.120) |
| Log(pool size) | 0.619*** | 3.16*** |
|  | (0.140) | (0.290) |
| ETH return (1d) | -1.71 | -0.623 |
|  | (4.71) | (3.80) |
| ETH return (7d) | 2.70 | -1.41 |
|  | (2.05) | (2.51) |
| ETH SD (30d) | 0.701 | 60.81*** |
|  | (19.02) | (15.73) |
| Constant | -2.23*** | -15.59*** |
|  | (0.500) | (1.62) |
|  |  |  |
| Observations | 351 | 426 |
| Adj R-squared | 0.108 | 0.568 |

Panel D: Unidentified contract

| VARIABLES | (1) cETH | (2) cWBTC | (3) cDAI | (4) cUSDC | (5) cUSDT |
| --- | --- | --- | --- | --- | --- |
| Lagged borrow rate | 1.726* | 0.086 | -0.118 | 0.374 | 0.566*** |
|  | (1.00) | (0.24) | (0.24) | (0.45) | (0.20) |
| Post COMP distribution | 3.21 | -2.97 | -5.26*** | -2.90 | 5.07** |
|  | (2.55) | (2.01) | (2.00) | (2.53) | (2.55) |
| Post * lagged borrow rate | -2.481** | -0.009 | 0.396 | -0.260 | -0.543** |
| | (1.10) | (0.35) | (0.25) | (0.47) | (0.21) |
| Lagged borrow reward | 0.052 | 0.021 | -0.065 | 0.031 | 0.288* |
| | (0.060) | (0.080) | (0.120) | (0.160) | (0.150) |
| Log(pool size) | 0.490*** | 0.740*** | 1.94*** | 2.89*** | 0.317 |
| | (0.170) | (0.190) | (0.340) | (0.300) | (0.200) |
| ETH return (1d) | 0.978 | -4.48 | 4.14 | 1.78 | -2.19 |
| | (3.59) | (5.80) | (4.34) | (5.12) | (6.24) |
| ETH return (7d) | -2.09 | 1.32 | -0.231 | 1.47 | -0.437 |
| | (1.48) | (2.84) | (2.15) | (2.57) | (3.63) |
| ETH SD (30d) | 34.62** | -3.63 | 21.75 | 36.46** | -68.85*** |
| | (16.69) | (27.56) | (15.63) | (16.76) | (18.08) |
| Constant | -6.09*** | 1.76 | 0.935 | -9.80*** | 0.735 |
| | (2.35) | (1.82) | (1.52) | (2.67) | (1.91) |
| | | | | | |
| Observations | 426 | 351 | 426 | 426 | 425 |
| Adj R-squared | 0.109 | 0.024 | 0.188 | 0.318 | 0.061 |

## APPENDIX: Compound Leaderboards

### Table A3: cToken holder leaderboard.
This table reports the cToken holders at different points in time. Historical holders are computed based on the entire history of the cToken transfers, while the holders as of February 2022 are obtained from Etherscan.io. Addresses of the top 10 holders are manually examined and classified using the scheme described in the Appendix.

Panel A: cETH holders

| cETH as of June 30, 2021 | Share (%) |
|---|---|
| [Celsius Network] 0x8aceab8167c80cb8b3de7fa6228b889bb1130ee8 | 13.77 |
| [Address] 0x716034c25d9fb4b38c837afe417b7f2b9af3e9ae | 12.41 |
| [Yield Agg / Vesper] 0xffc4c270244f9c0890c744f042f5f25f9ff8d4b5 | 4.99 |
| [Address] 0xc33d98e88682c883fe32b8f6620660692092d39f | 3.94 |
| [Address] 0x388b93c535b5c3ccdb14770516d7caf5590ed009 | 3.52 |
| [Address] 0x4740fa6b32c5b41ebbf631fe1af41e6fff6e2388 | 3.14 |
| [Gnosis Safe] 0xbc79855178842fdba0c353494895deef509e26bb | 3.00 |
| [AssetMgmt / Index Coop] 0xaa6e8127831c9de45ae56bb1b0d4d4da6e5665bd | 2.97 |
| [Address] 0x2baba0cba8241fda56871589835e0b05ec64ca41 | 2.05 |
| [Address] 0xc26b5977c42c4fa2dd41750f8658f6bd2b67869c | 1.80 |
| Sum of top 10 addresses | 51.60 |

| cETH as of Feb 11, 2022 | Share (%) |
|---|---|
| [Celsius Network] 0x8aceab8167c80cb8b3de7fa6228b889bb1130ee8 | 21.02 |
| [Address] 0x716034c25d9fb4b38c837afe417b7f2b9af3e9ae | 11.37 |
| [AssetMgmt / Index Coop] 0xaa6e8127831c9de45ae56bb1b0d4d4da6e5665bd | 4.66 |
| [InstaDApp] 0xfa5dcf356a2d80cf0c89d64a18a742edaf8d30e8 | 4.23 |
| [InstaDApp] 0x3a0dc3fc4b84e2427ced214c9ce858ea218e97d9 | 3.60 |
| [Address] 0xc26b5977c42c4fa2dd41750f8658f6bd2b67869c | 2.54 |
| [Holdnuat] 0x99fd1378ca799ed6772fe7bcdc9b30b389518962 | 2.41 |
| [Address] 0xbebcf4b70935f029697f39f66f4e5cea315128c3 | 2.37 |
| [Address] 0x1f244e040713b4139b4d98890db0d2d7d6468de4 | 1.89 |
| [Gnosis Safe] 0xe84a061897afc2e7ff5fb7e3686717c528617487 | 1.79 |
| Sum of top 10 addresses | 55.88 |

Panel B: cDAI holders

| cDAI as of June 30, 2021 | Share |
|---|---|
| [Address] 0x9b4772e59385ec732bccb06018e318b7b3477459 | 23.11 |
| [InstaDApp] 0x4c81ac8a069122d2a7146b08818fbaddcb2ff1f0 | 10.01 |
| [InstaDApp] 0xe4bed3988b25eb625466102f2d0bea1c9fafcd86 | 9.05 |
| [InstaDApp] 0x742fb193517619eecd6595ff106fce2f45488ebf | 5.79 |
| [InstaDApp] 0x9fe9dc57bf733bdafd0d6d4610d2d671f8dc974f | 4.38 |
| [YieldAgg / Idle] 0x78751b12da02728f467a44eac40f5cbc16bd7934 | 4.16 |
| [InstaDApp] 0x2cc308d515a73690ba58ed637d1b20b4b7324fcd | 3.71 |
| [Contract] 0x2cc308d515a73690ba58ed637d1b20b4b7324fcd | 3.71 |
| [InstaDApp] 0x10d88638be3c26f3a47d861b8b5641508501035d | 3.61 |
| [Address] 0x10bf1dcb5ab7860bab1c3320163c6dddf8dcc0e4 | 3.25 |
| Sum of top 10 addresses | 70.80 |

| cDAI as of Feb 11, 2022 | Share (%) |
|---|---|
| [Contract / Yearn] 0x1676055fe954ee6fc388f9096210e5ebe0a9070c | 21.81 |
| [InstaDApp] 0x1d1e63975486dfa6e7f28448ae224c9f41588642 | 16.87 |
| [InstaDApp] 0x10d88638be3c26f3a47d861b8b5641508501035d | 15.86 |
| [InstaDApp] 0x638e9ad05dbd35b1c19df3a4eaa0642a3b90a2ad | 14.86 |
| [InstaDApp] 0x41d207bc7e5d1f44aaf572d4a06cd0ef1ea2b01b | 6.45 |
| [Lending / Notional] 0x1344a36a1b56144c3bc62e7757377d288fde0369 | 6.25 |
| [DEX / Curve.fi] 0xa2b47e3d5c44877cca798226b7b8118f9bfb7a56 | 2.47 |
| [Contract / Bridge] 0x4aa42145aa6ebf72e164c9bbc74fbd3788045016 | 2.12 |
| [Address] 0x10bf1dcb5ab7860bab1c3320163c6dddf8dcc0e4 | 1.69 |
| [Fei Protocol: PCV] 0xe0f73b8d76d2ad33492f995af218b03564b8ce20 | 1.04 |
| Sum of top 10 addresses | 89.42 |

Panel C: cUSDC holders

| cUSDC as of June 30, 2021 | Share (%) |
|---|---|
| [Address] 0xb3bd459e0598dde1fe84b1d0a1430be175b5d5be | 6.16 |
| [YieldAgg / PoolTogether] 0xde9ec95d7708b8319ccca4b8bc92c0a3b70bf416 | 4.44 |
| [DEX / Curve.fi] 0xa2b47e3d5c44877cca798226b7b8118f9bfb7a56 | 4.42 |
| [Contract] 0x25a033316752ac9e443a10be07fa56c125b93c29 | 4.36 |
| [YieldAgg / Vesper] 0xbf84b97beabc953a7a2ad630940065b69d24c912 | 4.30 |
| [Contract] 0xdaa037f99d168b552c0c61b7fb64cf7819d78310 | 4.22 |
| [YieldAgg / Idle] 0x5274891bec421b39d23760c04a6755ecb444797c | 3.40 |
| [Address] 0xd31ab5ab8cd0f482f5728888519b5c39b5a4a6a0 | 3.07 |
| [Address] 0x22fa8cc33a42320385cbd3690ed60a021891cb32 | 2.92 |
| [Address] 0x1e17f8876b175d37ebe08849434973c051261461 | 2.78 |
| Sum of top 10 addresses | 40.07 |

| cUSDC as of Feb 11, 2022 | Share (%) |
|---|---|
| [YieldAgg / Yearn] 0x342491c093a640c7c2347c4ffa7d8b9cbc84d1eb | 7.25 |
| [InstaDApp] 0x3a0dc3fc4b84e2427ced214c9ce858ea218e97d9 | 6.62 |
| [Address] 0xabde2f02fe84e083e1920471b54c3612456365ef | 6.50 |
| [YieldAgg / Angle] 0x6d7ccd6d3e4948579891f90e98c1bb09a8c677ea | 6.33 |
| [Lending / Notional] 0x1344a36a1b56144c3bc62e7757377d288fde0369 | 5.91 |
| [Address] 0xdb7030beb1c07668aa49ea32fbe0282fe8e9d12f | 4.47 |
| [YieldAgg / Yearn] 0x7900c70a377f89df29d1d1939469ae3b74c5b740 | 4.45 |
| [Address] 0xb3bd459e0598dde1fe84b1d0a1430be175b5d5be | 4.07 |
| [Justin Sun] 0x3ddfa8ec3052539b6c9549f12cea2c295cff5296 | 3.70 |
| [Proxy Contract] 0x2d15fcd5d6849a72f0bda676a1f2f1aede7467f5 | 3.64 |
| Sum of top 10 addresses | 52.94 |

Panel D: cUSDT holders

| cUSDT as of June 30, 2021 | Share (%) |
|---|---|
| [Address] 0xf23913349c935dacc7c51ee692961ebc0d69fc35 | 7.25 |
| [Address] 0x7d6149ad9a573a6e2ca6ebf7d4897c1b766841b4 | 4.69 |
| [Address] 0xb99cc7e10fe0acc68c50c7829f473d81e23249cc | 4.59 |
| [Address] 0x102fa4db3bc6a70d85513a4f424e739ef922bf1e | 3.15 |
| [Address] 0x0c731fb0d03211dd32a456370ad2ec3ffad46520 | 2.84 |
| [Address] 0x01d2e7cea783b0458ba6b58e93906b19b9741889 | 2.65 |
| [Address] 0x133b590c0d9d9051c78f959ac1eb435c7676dcdc | 2.63 |
| [Address] 0xfb849f0b58ee2421b788005c15c1485f384de73f | 2.07 |
| [Address] 0xd5433168ed0b1f7714819646606db509d9d8ec1f | 1.86 |
| [Address] 0x251b32806b4cd6bc50470ea94a07462609ef798d | 1.67 |
| Sum of top 10 addresses | 33.41 |

| cUSDT as of Feb 11, 2022 | Share (%) |
|---|---|
| [Justin Sun] 0x3ddfa8ec3052539b6c9549f12cea2c295cff5296 | 9.81 |
| [Address] 0x1a8c53147e7b61c015159723408762fc60a34d17 | 8.34 |
| [Address] 0xb99cc7e10fe0acc68c50c7829f473d81e23249cc | 6.40 |
| [Alameda Research] 0x712d0f306956a6a4b4f9319ad9b9de48c5345996 | 6.14 |
| [Address] 0xe9bf81b432e5bf34995afae08747c530c8406c4d | 4.32 |
| [Address] 0x01d2e7cea783b0458ba6b58e93906b19b9741889 | 2.03 |
| [Address] 0x638edf6438ec145454b3ea483fea7339377fe80f | 1.91 |
| [FTX Exchange] 0x2faf487a4414fe77e2327f0bf4ae2a264a776ad2 | 1.62 |
| [Address] 0xfb849f0b58ee2421b788005c15c1485f384de73f | 1.60 |
| [Address] 0xf07766108cdb54082f7b06cad20d6adab1342d46 | 1.42 |
| Sum of top 10 addresses | 43.58 |

**Table A4: Borrower leaderboard.**
This table reports the cumulative loans drawn from cToken contracts until June 2021. Net deposits are the net dollar amount that are deposited or withdrawn from cToken contracts. If net deposits are negative, the address owner likely acquired cTokens via other means and use them to redeem for deposits from cToken contracts. Depending on the secondary market price of the cTokens, buying them to redeem could provide arbitrage profit. COMP governance tokens claimed are reported in units.

| Address | Total Loans (USD m) | Net Deposits (USD m) | COMP Claimed (units) |
|---|---|---|---|
| [Address] 0x4740fa6b32c5b41ebbf631fe1af41e6fff6e2388 | 4,331.17 | 81.5 | 196.4 |
| [Address] 0x2bdded18e2ca464355091266b7616956944ee7ee | 3,781.70 | -407.4 | 50,349.2 |
| [yDai exploiter] 0x62494b3ed9663334e57f23532155ea0575c487c5 | 2,940.41 | 0.0 | 0.0 |
| [InstaDApp] 0x691d4172331a11912c6d0e6d1a002e3d7ced6a66 | 1,997.00 | 0.0 | 0.0 |
| [Contract / Yearn] 0x77b7cd137dd9d94e7056f78308d7f65d2ce68910 | 1,788.63 | -3.1 | 6,659.8 |
| [Alameda Research] 0x4deb3edd991cfd2fcdaa6dcfe5f1743f6e7d16a6 | 1,440.20 | -6,693.5 | 6,225.3 |
| [Contract / Yearn] 0x4031afd3b0f71bace9181e554a9e680ee4abe7df | 1,389.26 | -6.8 | 15,331.9 |
| [Three Arrows Capital] 0x3ba21b6477f48273f41d241aa3722ffb9e07e247 | 1,357.27 | -130.2 | 27,088.7 |
| [Address] 0x3aa39dff4964043a61d94029fcedaac2f02f3187 | 1,162.90 | -22.0 | 4,741.5 |
| [Contract / Yearn] 0xe68a8565b4f837bda10e2e917bfaaa562e1cd143 | 1,075.70 | -3.4 | 7,215.2 |
| [Proxy Contract] 0xccb06b8026cb33ee501476af87d5ccaf56883112 | 1,055.34 | -162.5 | 61,569.4 |
| [Contract / Yearn] 0x339dc96a37dba86008126b3391db77af93cc0bd9 | 956.85 | -0.3 | 1,893.3 |
| [Address] 0x9b4772e59385ec732bccb06018e318b7b3477459 | 955.91 | 1,126.8 | 4,188.3 |
| [Address] 0x1f99aaa8b4fb631d25b38b8a9099ef8f2611e46b | 908.00 | 23.4 | 4,708.1 |
| [InstaDApp] 0x2a1739d7f07d40e76852ca8f0d82275aa087992f | 813.70 | 0.0 | 0.0 |
| [Contract / Yearn] 0x55ec3771376b6e1e4ca88d0eea5e42a448f51c7f | 793.60 | -0.4 | 2,151.3 |
| [Contract / Yearn] 0x4d7d4485fd600c61d840ccbec328bfd76a050f87 | 792.51 | -3.7 | 10,108.7 |
| [Address] 0x767ecb395def19ab8d1b2fcc89b3ddfbed28fd6b | 767.72 | -3,585.8 | 8,687.9 |
| [Address] 0xb1adceddb2941033a090dd166a462fe1c2029484 | 761.99 | -3,085.0 | 6,119.4 |
| [InstaDApp] 0x06cb7c24990cbe6b9f99982f975f9147c000fec6 | 711.26 | -1,813.1 | 47.0 |